\documentclass[preprint,superscriptaddress,amsmath,amssymb,amsfonts,aps,prapplied,floatfix,onecolumn,tightenlines]{revtex4-1}

\usepackage{graphicx}
\usepackage{color,xcolor}
\graphicspath{{Figures/}}
\usepackage{bm}
\usepackage{textcomp}

\usepackage{float}
\usepackage[normalem]{ulem}
\usepackage{subcaption}

\newcommand{\blue}{\color{black}}    
\newcommand{\red}{\color{black}}      
\newcommand{\vio}{\color{black}}

\newcommand{\brown}{\color{black}}

\begin{document}


\title{Structure and electrical behavior of silicon nanowires prepared by MACE process }

\author{R. Plugaru}
\thanks{These two authors contributed equally}
\affiliation{National Institute for Research and Development in Microtechnologies- IMT Bucharest,
077190 Voluntari, Romania}

\author{E. Fakhri}
\thanks{These two authors contributed equally}
\affiliation{Department of Engineering, Reykjavik University, Menntavegur 1, IS-102 Reykjavik, Iceland}

\author{C. Romanitan}
\affiliation{National Institute for Research and Development in Microtechnologies- IMT Bucharest,
077190 Voluntari, Romania}

\author{I. Mihalache}
\affiliation{National Institute for Research and Development in Microtechnologies- IMT Bucharest,
077190 Voluntari, Romania}

\author{G. Craciun}
\affiliation{National Institute for Research and Development in Microtechnologies- IMT Bucharest,
077190 Voluntari, Romania}

\author{N. Plugaru}
\affiliation{National Institute for Research and Development in Microtechnologies- IMT Bucharest,
077190 Voluntari, Romania}

\author{H. \"O. \'Arnason}
\affiliation{Department of Engineering, Reykjavik University, Menntavegur 1, IS-102 Reykjavik, Iceland}

\author{M. T. Sultan}
\affiliation{Science Institute, University of Iceland, Dunhaga 3, 107 Reykjavik, Iceland}

\author{G. A. Nemnes}
\affiliation{Horia Hulubei National Institute for Physics and Nuclear Engineering, 077126 Magurele-Ilfov, Romania}
\affiliation{University of Bucharest, Department of Physics, 077125 Magurele-Ilfov, Romania}

\author{S. Ingvarsson}
\affiliation{Science Institute, University of Iceland, Dunhaga 3, 107 Reykjavik, Iceland}

\author{H. G. Svavarsson}
\affiliation{Department of Engineering, Reykjavik University, Menntavegur 1, IS-102 Reykjavik, Iceland}

\author{A. Manolescu}
\affiliation{Department of Engineering, Reykjavik University, Menntavegur 1, IS-102 Reykjavik, Iceland}


\begin{abstract}
We report on the structure and electrical characteristics of silicon nanowire arrays prepared by metal assisted chemical etching (MACE) method, investigated by cross-sectional scanning electron microscopy (SEM) and high resolution X-ray diffraction (HR-XRD) methods. SEM micrographs show arrays of merged parallel nanowires, with lengths of 700 nm and 1000 nm, resulted after 1.5 min and 5 min etching time, respectively. X-ray reciprocal space maps (RSMs) around Si (004) reciprocal lattice point indicate the presence of 0D structural defects rather than of extended defects. The photoluminescence spectra exhibit emission bands at 1.70 eV and 1.61 eV, with intensity significantly higher in the case of longer wires and associated with the more defected surface. The transient photoluminescence spectroscopy reveals average lifetime of 60 \textmu s and 111 \textmu s for the two SiNW arrays, which correlate with a larger density of defects states in the latest case. The I-V characteristics of the nanowires, show a memristive behavior with the applied voltage sweep rate in the range 5V/s - 0.32V/s. We attribute this behavior to trap states which control the carrier concentration, and model this effect using an equivalent circuit. Photogeneration processes under excitation wavelengths in visible domain, 405 nm - 650 nm, and under light intensity in the range 20 - 100 mW/cm$^2$ provided a further insight into the trap states. 
\end{abstract}





\maketitle

\section{Introduction}
Silicon nanowire (SiNW) arrays with controlled morphology (porosity, length, orientation) have been efficiently prepared by metal assisted chemical etching (MACE) processing \citep{chien2020high,han2014metal,azeredo2013silicon}, aiming to widen their applicability to performant light emitting devices, photodetectors, energy storage and conversion, or sensors \citep{baraban2019hybrid,guan2021single,Zaibi2022}. Yet, defective surfaces of SiNWs resulting from chemical-assisted preparation processes can affect the electric parameters of the devices \citep{moh2013effect,dan2011dramatic}. 

{\red The dynamic hysteresis of the electrical characteristics of a nano-electronic device when the applied voltage changes in time is one of the first indications of the presence of a charge trapping mechanism inside the devices with a relaxation time comparable to the time of the voltage variation. This phenomenon is used in memristive device, often based on metallic nanowires \cite{Milano19}, but also on silicon nanowires \cite{carrara2012memristive,Sacchetto14}. Sensing devices based on SiNWs have also been proposed, where the dynamic occupation of the nanowire surface states created by the external charges from the adsorbed biomolecules modify the hysteresis loop  \cite{puppo2016surface}. Or field effect transistors where the nanowires are in contact with a dielectric, or polymer  \cite{rajeev2017effect}.}

{\red The electrical hysteresis is also present in solar cells based on perovskite materials, and associated to the degradation of the cell, due to ionic migration, charge accumulation at interfaces, and their influence on the photogenerated current \citep{seki2016equivalent,chen2015impact,tress2016inverted}.  
Apart of the possible hysteretic effects, the surface states of SiNWs are also important for the characteristics of nanostructured solar cells, where the electrons trapped by surface states can act as a gate bias that enhances the photoconductivity \citep{moiz2020design,yu2016design,rurali2010colloquium}. 
}


In this work we report on the electrical response of Al/SiNWs/Al device structures with pristine SiNW arrays prepared by the MACE method. {\red The nanowires are highly imperfect and in lateral contact to each other, forming a system of inter-connected wires rather than independent wires.} The current-voltage (I-V) and capacitance-voltage (C-V) curves, measured in dark and under various illumination conditions in terms of wavelengths and intensity, are analyzed using an equivalent circuit with lump elements. The model, although simplistic, is able to reproduce satisfactorily the normal and inverted hysteresis observed in the I-V curves. We suggest that the intimate mechanism accounting for the hysteresis may be related to the effect of charge carrier trapping and detrapping at the surface states present in the pristine SiNW arrays. We show that traps filling by tuning the light wavelength may be used as a method to determine their origin and electronic properties.

\section{Experimental}

\begin{figure}
    \centering
    \includegraphics[width=0.7\linewidth]{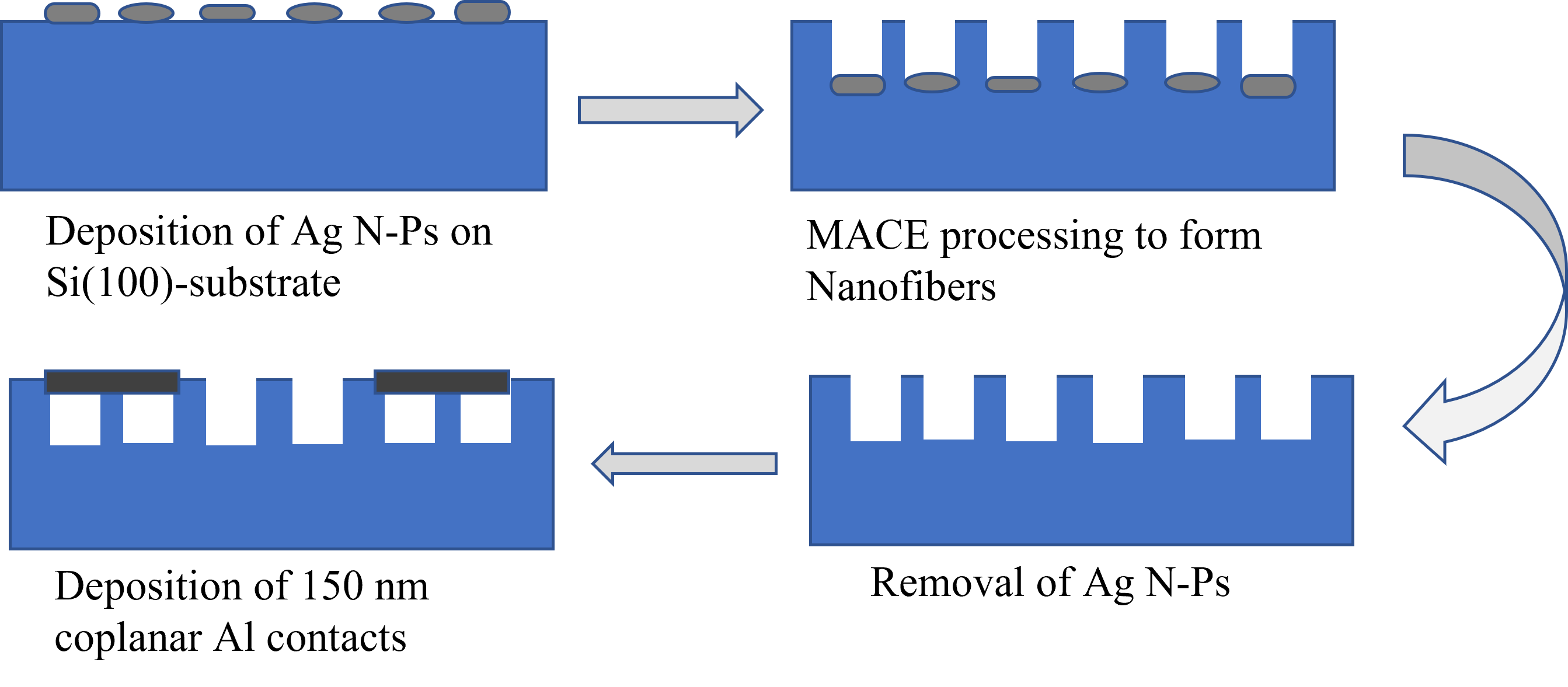}
    \caption{Schematic illustration of the MACE steps for SiNWs preparation and Al/SiNWs/Al structures fabrication.} 
    \label{schematic}
\end{figure}

Inter-connected silicon nanowire arrays have been fabricated via metal-assisted chemical etching (MACE) as shown in the flow chart presented in Fig.\ \ref{schematic} Single-sided polished samples ($1\times1$~cm$^2$) cut from (100) oriented silicon wafers (p-type, with resistivity of $1-10~\Omega\cdot$cm and thickness of $625$~µm) were used in the process.
The samples were cleaned with acetone, methanol, isopropanol, and deionized water followed by drying in N$_2$ gas flow, then treated with HF:H$_2$O (1:3) solution for 3 min to remove the native oxide. To obtain the SiNW arrays, the polished side of the samples were coated with Ag nanoparticles by immersing them in HF [3~M]:AgNO$_3$ [1.5~mM] solution for 1 min. The coated samples were then cleaned with DI-water to remove the excess of Ag nanoparticles. The samples subsequently underwent etching process by immersing them in an etching solution of HF [5~M]:H$_2$O$_2$[0.4~M] for 1.5 min and 5 min, respectively, followed by immersion in H$_2$O:HNO$_3$ (3:1) to remove residual Ag nanoparticles. Finally, the samples were cleaned with DI-water, then dried with N$_2$ gas. For electrical measurements, two coplanar Al contacts, $2\times$10~mm$^2$ each, with the thickness of 150~nm were deposited on the surface of the samples via a hard mask. The distance between the two contacts is 6~mm. Contacts deposition was made by using an electron beam evaporator (Polyteknik Cryofox Explorer 600 LT).

Micrographs of the SiNWs were recorded on the top and at a tilt angle to observe the in-depths of the structures, by using a field emission scanning electron microscope (FE-SEM), FEI NovaTM NanoSEM 630. The structural characteristics of the SiNW arrays were investigated by using a SmartLab X-ray diffraction system from Rigaku Corp. (Osaka, Japan). X-ray reciprocal space maps (RSMs) around Si (004) reciprocal lattice point were recorded in triple-axis configuration (ultra-high resolution) with a four-bounce Ge monochromator with two reflections at incidence and a two-bounce Ge monochromator with two reflections in the front of detector. Bending profiles were obtained using grazing-incidence XRD in asymmetric skew geometry on (111) reflection. In this configuration, the incidence angle of the source was varied from $0.5^{\circ}$ to $4^{\circ}$ to obtain different X-ray penetration depth. 

The photoluminescence (PL) emission spectra of the SiNW arrays were recorded with an Edinburgh FL920 fluorescence spectrophotometer equipped with microsecond flashlamp as an excitation source. Time-correlated single photon counting (TCSPC) technique was used to determine the photoluminescence lifetime, using excitation at 300~nm and recording the emission at 770~nm wavelength. FAST Version 3.4.2. Edinburgh Instruments Ltd software was used for experimental fit.

The I-V and C-V characteristics were measured using a  Keithley 2400 and SCS 4200 Keithley system, in dark and under light illumination with various wavelengths, as well as under white light using a solar simulation lamp, with intensities in the range 20-100 mW/cm$^2$. 
The curves were recorded by forward and reverse sweeping the applied voltage in the range -10V to +10V.

\section{Results}
\subsection{Structure and PL properties of silicon nanowires}
\begin{figure}
    \centering
    \includegraphics[width=0.7\linewidth]{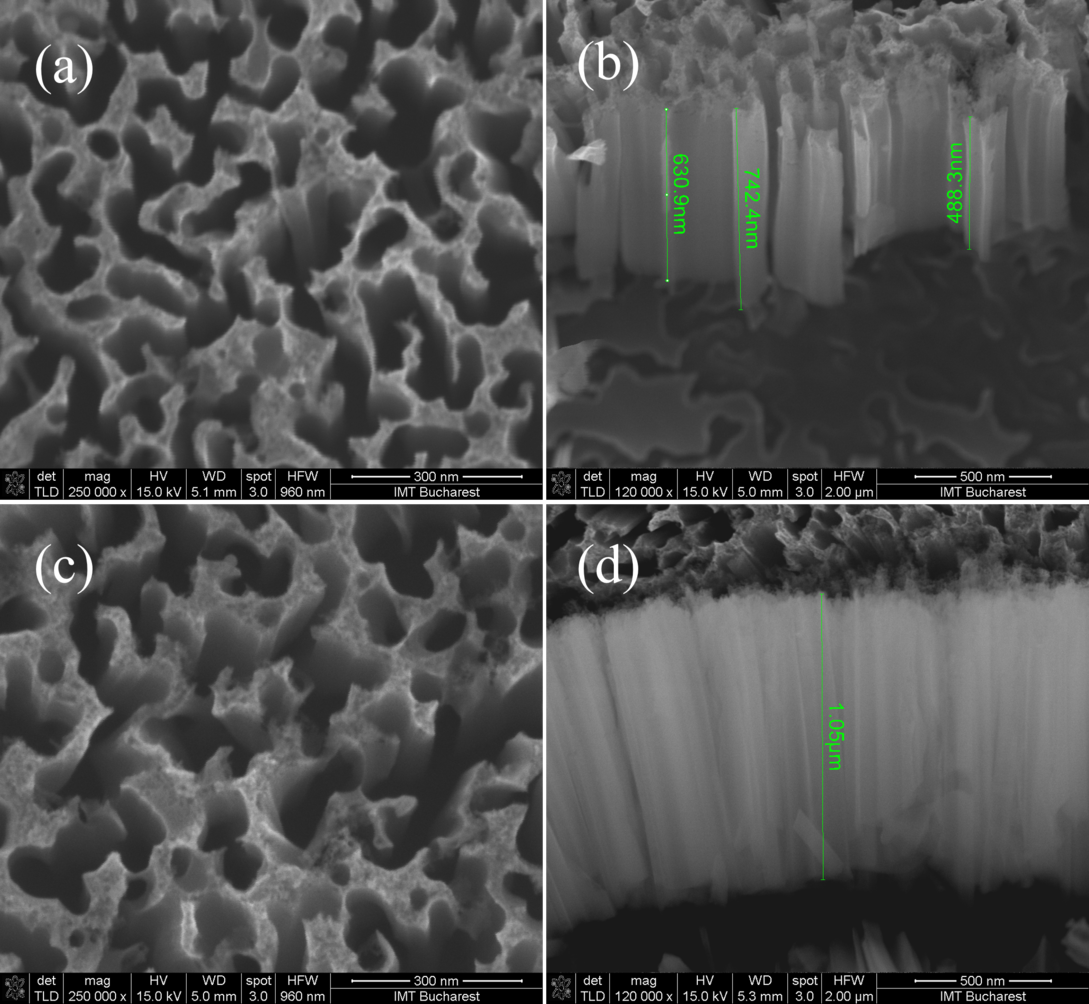}
    \caption{SEM micrographs of the SiNWs arrays prepared by a MACE process. The etching time was: (a), (b) 1.5 min and (c), (d) 5 min.} 
    \label{Figure 2}
\end{figure}

\begin{figure}
    \centering
    \includegraphics[width=0.7\linewidth]{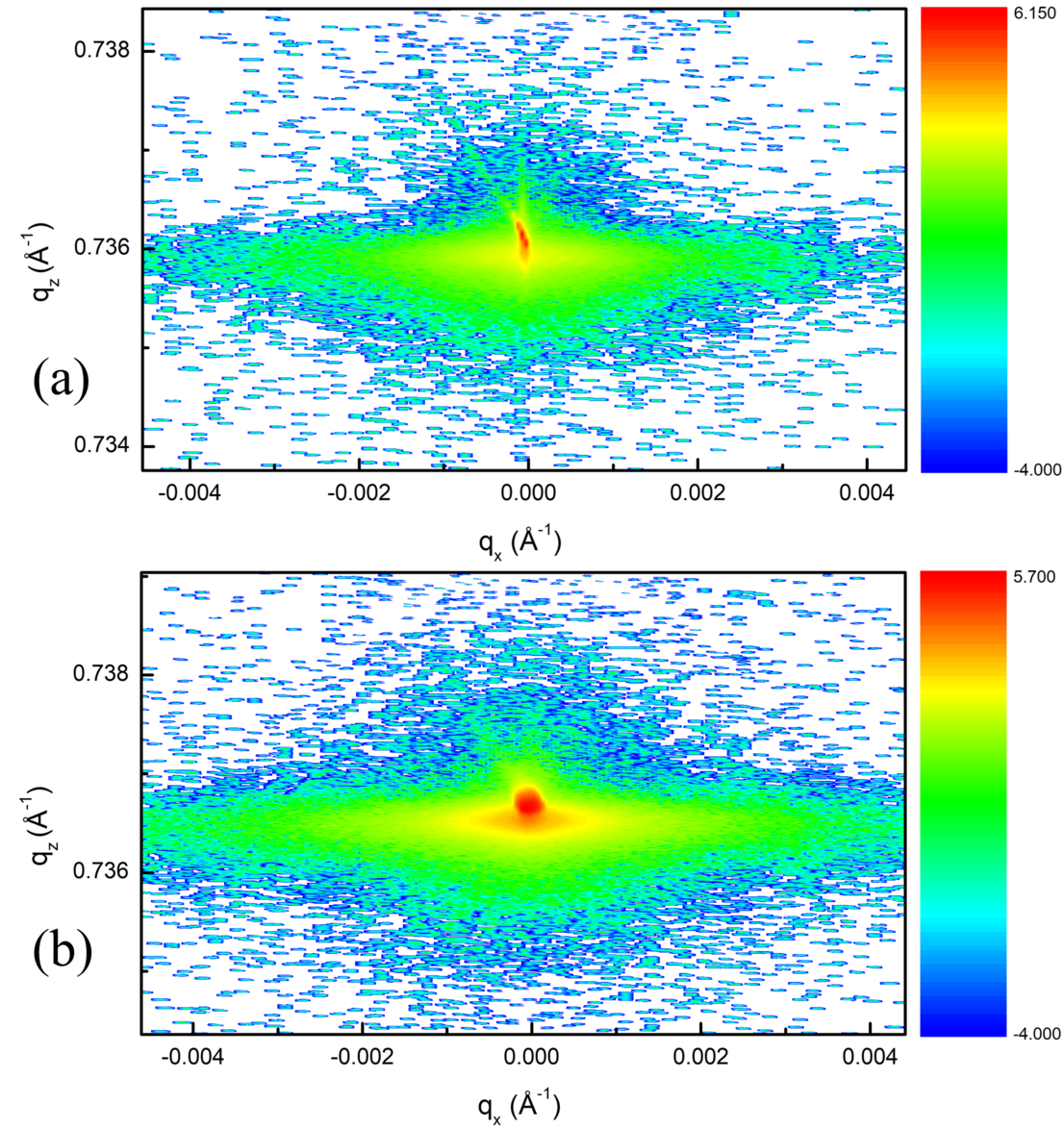}
    \caption{X-ray reciprocal space maps (RSMs) recorded near Si (004) reciprocal lattice point on the nanowire arrays with length of (a) 700 nm and (b) 1000 nm SiNWs.} 
    \label{Figure 3}
\end{figure}

The top and cross-sectional SEM micrographs of the SiNW arrays obtained at 1.5 min and 5 min etching time, are shown in Fig.\ \ref{Figure 2}a-d.
The length of the NWs is about 700 nm after 1.5 min etching (see Figs.\ \ref{Figure 2}a-b) and 1000 nm after 5 min etching (Figs.\ \ref{Figure 2} c-d). In the following we refer to {\vio these two samples as SiNW$_{\mathrm {short}}$ and SiNW$_{\mathrm {\mathrm {long}}}$, respectively.} 
\textcolor{black}{There is a saturation value so that the lengths of the wires are limited with the respect of the concentrations of the oxidizing agent H$_2$O$_2$,  AgNO$_3$ concentration, and specific resistivity ($\rho$) of bulk Si \citep{fakhri2021synthesis}}.
As seen, the wires are laterally interconnected and form continuous structures of walls in both arrays. According to ImageJ analysis, the coverage area is 36\% and 32\% for {\vio SiNW$_{\mathrm {short}}$ and SiNW$_{\mathrm {long}}$}, respectively, which suggests that the longer etching time leads to a higher porosity. \textcolor{black}{Note that the coverage area means the surface occupied by tips of SiNWs observed in SEM top view image and by longer etching time the diameter of SiNWs decreases and as result the coverage area decreases.}
In order to investigate the microstructural features of the nanowire arrays, we performed X-ray diffraction in high resolution setup. X-ray reciprocal space maps (RSMs) around Si (004) reciprocal lattice point give information regarding the out-of-plane lattice value, relative lattice strain and the crystal imperfections. X-ray RSMs along $(q_z,q_x)$ coordinates for {\vio SiNW$_{\mathrm {short}}$ and SiNW$_{\mathrm {long}}$} arrays are presented in
Figs.\ \ref{Figure 3}a,b. The reciprocal space coordinates $q_x$ and $q_z$ are projections of the scattering vector along [100] and [001] directions, respectively, and are related with the angular coordinates as: $q_x=[2\sin (\omega-\theta)] / \lambda$ and $q_z=(2\sin \theta) / \lambda$.  

The X-ray reciprocal space maps present an intense peak located {\red around $q_z \in (0.7360-0.7365)$ \AA$^{-1}$}. Using the crystallographic relations for cubic crystals, the lattice constant $a$  can be expressed as $\frac{4}{q_z}$ \citep{stanchu2017asymmetrical}, giving a lattice constant equal to $5.43$ \AA,  which corresponds to the lattice parameter of bulk Si. This is an indication that the MACE process does not affect the value of the lattice parameter of the samples. Further, it can be observed that the spot broadening increases with increasing the nanowire length in both $q_z$ and $q_x$ direction. The broadening of the RSM spot can be ascribed to the occurrence of bending and torsion of nanowire array, which is more pronounced for the longer (1000~nm) SiNWs, due to a higher surface energy. At the same time, the X-ray scattering in the reciprocal space looks different. For instance, the area elongated along $q_x$, which is related to the diffuse scattering, could be determined by crystal imperfections (e.g. point defects, extended defects, or stacking faults), and broader angular dispersion is observed for the longer nanowires. 
The cross section of intensity distribution of our RSM along $q_x$ is presented below in Fig.\ \ref{Figure 4}a.

\begin{figure}
    \centering
    \includegraphics[width=0.7\linewidth]{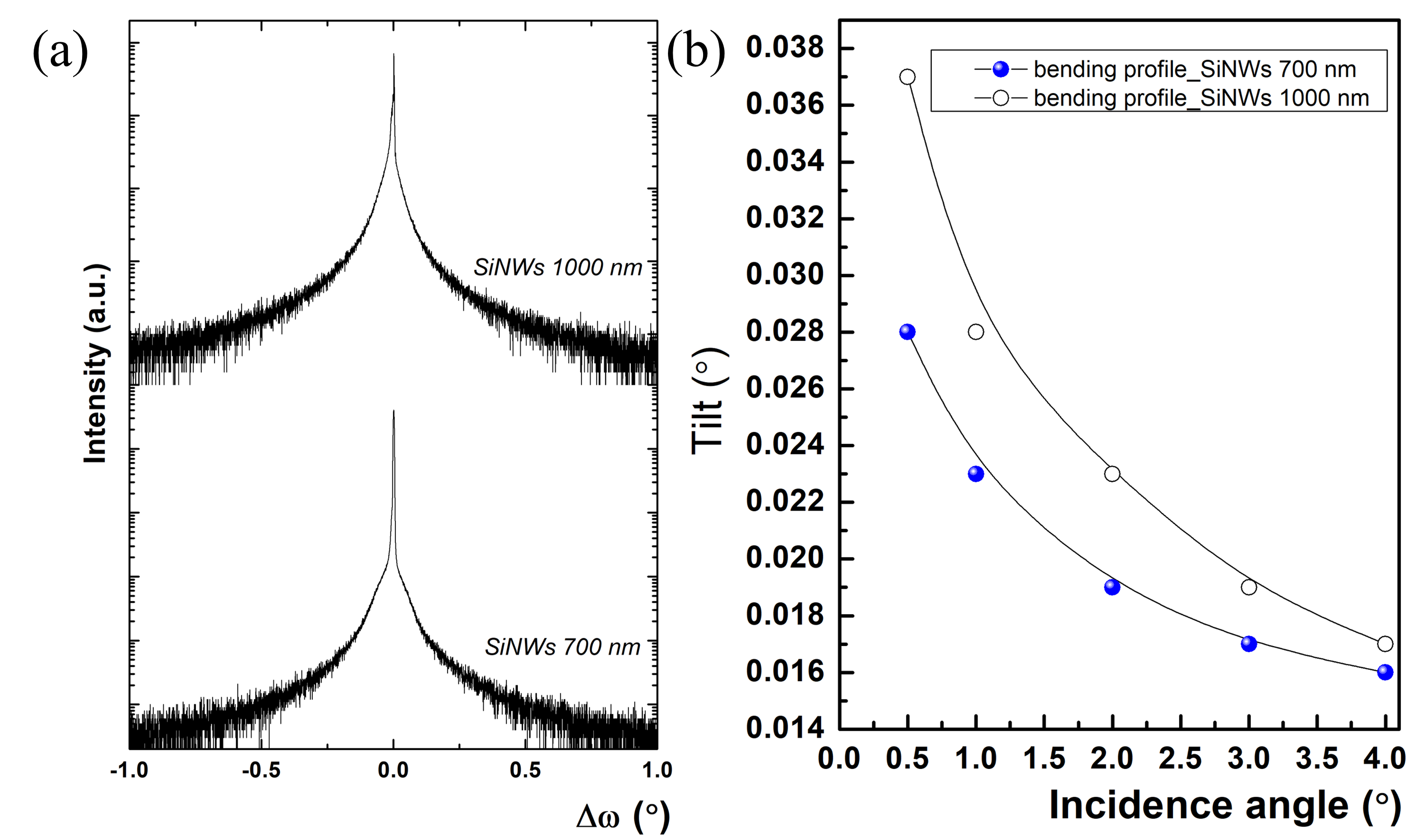}
    \caption{Cross section of intensity distribution along q$_x$  direction for short and long nanowire arrays (a) and (b) bending profiles.} 
    \label{Figure 4}
\end{figure}

The X-ray rocking curves indicate two types of scattering: (1) the narrow peak is related to the specular scattering ($I_{\mathrm {spec}}$), also called Bragg scattering, where the X-ray scattering has taken place on the atomic planes; (2) the broad feature indicates the presence of the X-ray diffuse scattering (XDRS), further denoted as $I_{\mathrm {diff}}$. This scattering is determined by the structural imperfections in the Si lattice. To have a qualitative description of the X-ray scattering, we obtain the ratio between ${I_{\mathrm{diff}}}/{(I_{\mathrm{diff}}+I_{\mathrm{spec}})}$. For instance, the {\vio shorter SiNWs exhibits a ratio of 0.31, whereas the longer SiNWs exhibits ratio of} 0.94. Clearly, this ratio can be viewed as a measure of the structural defect density in our samples. It is reasonable to assume that the longer SiNWs possesses a higher density of the structural defects, being promoted by the strain relaxation processes due to the bending and torsion phenomena \citep{kaganer2016elastic}.

To prove the existence or absence of the strain relaxation processes, we obtained bending profiles of our samples, which correspond to
the average tilt of the arrays - Figure \ \ref{Figure 4}b. These profiles were obtained in the framework of grazing-incidence X-ray diffraction on highly-asymmetric (111) reflection, which allowed us to attain different X-ray penetration depths, varying the incidence angles of the X-ray source. Further details regarding the grazing-incidence X-ray diffraction technique on (111) in nanowires, as well as for the bending profiles can be found in \citep{romanitan2019unravelling}.

The evolution of the FWHM of the X-ray spectra with the incidence angle gives the tilt of nanowire array at different penetration depths. One may observe that the shorter nanowires determine a smaller tilt, e.g. $0.028^{\circ}$, comparing to the longer ones which have an average tilt of $0.037^{\circ}$. It is clear that the higher tilt for the longer nanowires is determined by a higher surface energy of the nanowire array. However, for both samples the bending profiles do not present dips, whose occurrence can be assigned to a quasi-local manifestation of some relaxation mechanisms in the nanowires. The absence of the strain relaxation processes can be further attributed to the absence of the extended structural defects, such as edge and screw threading dislocations. This is {\red possible by} taking into account the small length of our arrays. Also, previous investigations in highly dense nanowire arrays prepared by MACE showed the occurrence of edge and screw threading dislocations only for {\vio wires longer than} 9~ \textmu m \citep{romanitan2019unravelling}. At the same time, we must consider the previous findings from the rocking curves profiles, indicating a relationship between the array length and the XRDS intensity, which was attributed to the presence of the structural defects. 

\begin{figure}
    \centering
    \includegraphics[width=0.7\linewidth]{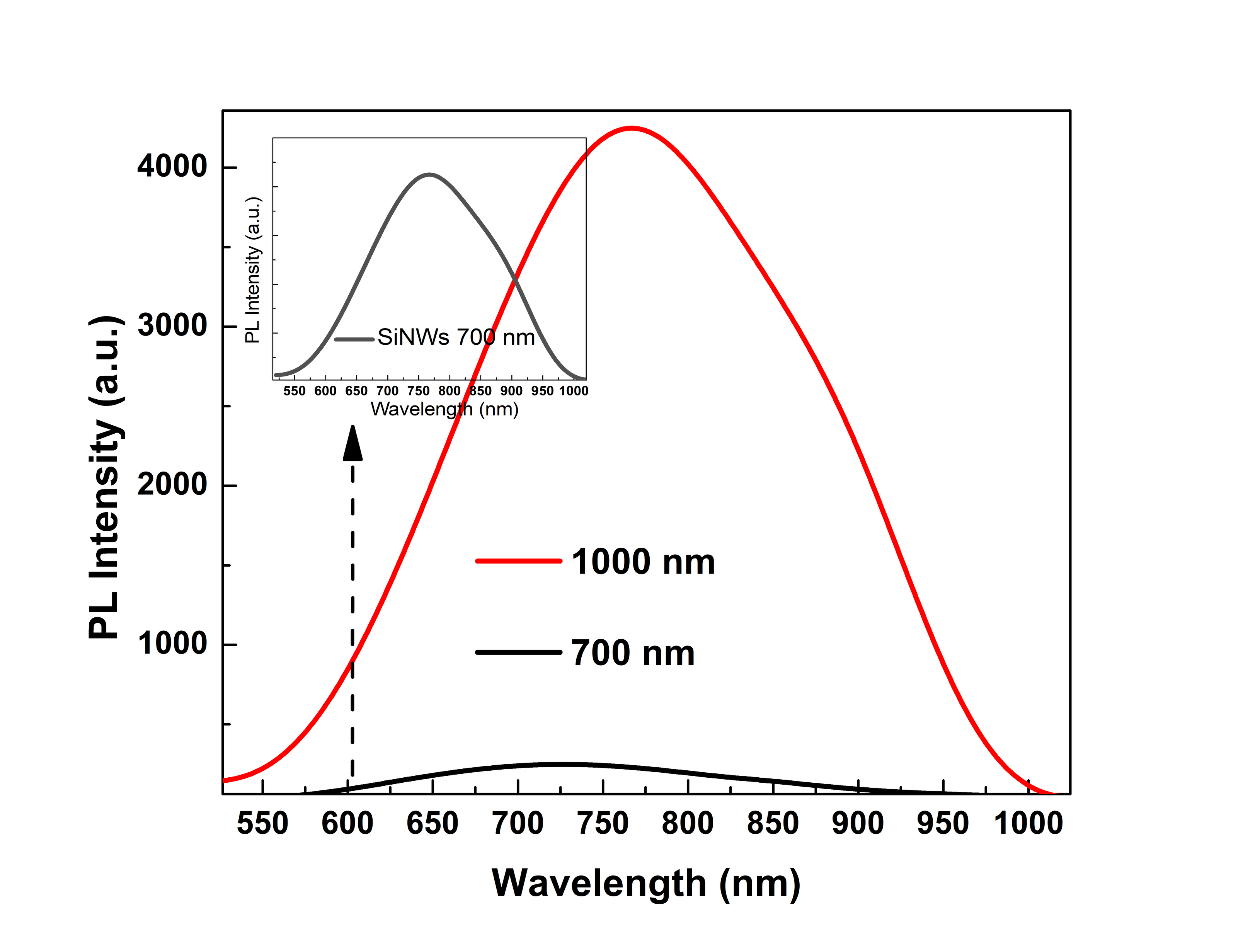}
    \caption{PL spectra of SiNWs arrays under excitation with Xe lamp at wavelength of 350 nm and 450 W power. } 
    \label{Figure 5}
\end{figure}

The XRD findings indicate that the MACE process has determined the formation of 0D defects, nanocrystals or nanopores on the surface of SiNWs, rather than extended structural defects. The nature of structural defects is analyzed by recording the photoluminescence (PL) emission from SiNWs arrays. The PL spectra are shown in Fig.\ \ref{Figure 5}. 

The SiNW$_{\mathrm {short}}$ presents a rather weak PL intensity, about {\vio 1/15 of the PL for the} SiNW$_{\mathrm {long}}$. The spectrum is centered at 730 nm (1.70 eV). Sample SiNW$_{\mathrm {long}}$ shows an intense and broad PL emission spectra in the VIS-NIR region, centered at 770 nm (1.61 eV). The red-shift in energy of 80~meV of the PL maximum position, observed for the longer NWs, could be related to the structure of interconnected skeletons and increased porosity, as resulted during the MACE process with a longer duration \citep{fakhri2021synthesis}.
\textcolor{black}{The observed red-shift with longer etching time is in agreement with our previous results \citep{fakhri2021synthesis}. It has however worth noting that others researchers \citep{congli2013synthesis}, have observed blue-shift upon increasing etching time and correlated that with the presence of SiO$_{\rm x}$. It appears that in our case, the SiO$_{\rm x}$ fraction is not increasing with the etching time and consequently has a low contribution to the PL emission which strongly shifts towards NIR spectral range.} A slow (S) band in the red-yellow spectral range with long microsecond decay times is reported and analyzed in porous silicon nanostructures PL spectra and attributed to phonon-assisted exciton recombination within the silicon nanostructure \citep{canham2020introductory}. It was also reported that TEM images of the luminescent SiNWs prepared by MACE technique reveal that the surfaces of the SiNWs are very rough, with a few nano-sized silicon particles being attached to the SiNWs. The PL spectrum of such SiNWs was peaked at 700~nm for an excitation wavelength of 400~nm \citep{jung2019optical}. 

Lin et al. \citep{lin2010synthesis} reported that SiNWs synthesized via MACE exhibit a nanoporous structure. The PL emission band in the red region, at 730 nm was attributed to the excitons captured by the interface states between the Si nanostructures and the native oxide layer. The PL intensity increases with the porosity \citep{lin2010synthesis}. It was also reported that longer etching time, or higher H$_2$O$_2$ concentration could facilitate the diffusion and nucleation of Ag ions and effectively enhance the porosity of the nanowires \citep{razek2014vertically}. Recently, it was shown that MACE-produced SiNW arrays are covered with porous structures, silicon nanocrystals, which result from the lateral etching of NWs sidewalls. The broad PL spectrum centered at 695~nm (1.78~eV) is attributed to radiative recombination of excitons in these nanocrystals \citep{naffeti2020highly}. 
In our experiment, even if the length of the NWs does not differ very much, the PL intensity is substantially higher in the case of long wires which suggests the formation of a larger number of luminescence centers. 

The enhancement of the PL intensity and the wavelength red-shift could be attributed to enhanced porous structure of the SiNWs surface and also of a porous Si layer formed at the base of SiNWs, resulting after a longer MACE process \citep{leontis2013structure,okayama2009ordering}. \textcolor{black}{Previous studies reported that HF post etching treatments of SiNWs are mandatory in order to obtain light emission \citep{amri2020photoluminescence}.Other experiments demonstrated that H$_2$O$_2$ could favor PL emission, which is attributed to SiO$_2$ layer formation on the NWs surface \citep{congli2013synthesis}. However, in this work a different etching process in terms of reagents concentration leads to a significant intensity of the PL emission, without any post-treatment, as also observed by reference \citep{wang2012electrochemically}. This result underlines the essential role of the etching solution concentration on the formation of various light emitting centers, such as Si nanostructures, SiO$_2$ layer, other specific Si bonding structure. }
The PL lifetime of the SiNWs arrays was measured by transient photoluminescence spectroscopy (TCSPC) method.  \\

\vspace{-5mm}
\begin{figure}
    \centering
    \includegraphics[width=0.7\linewidth]{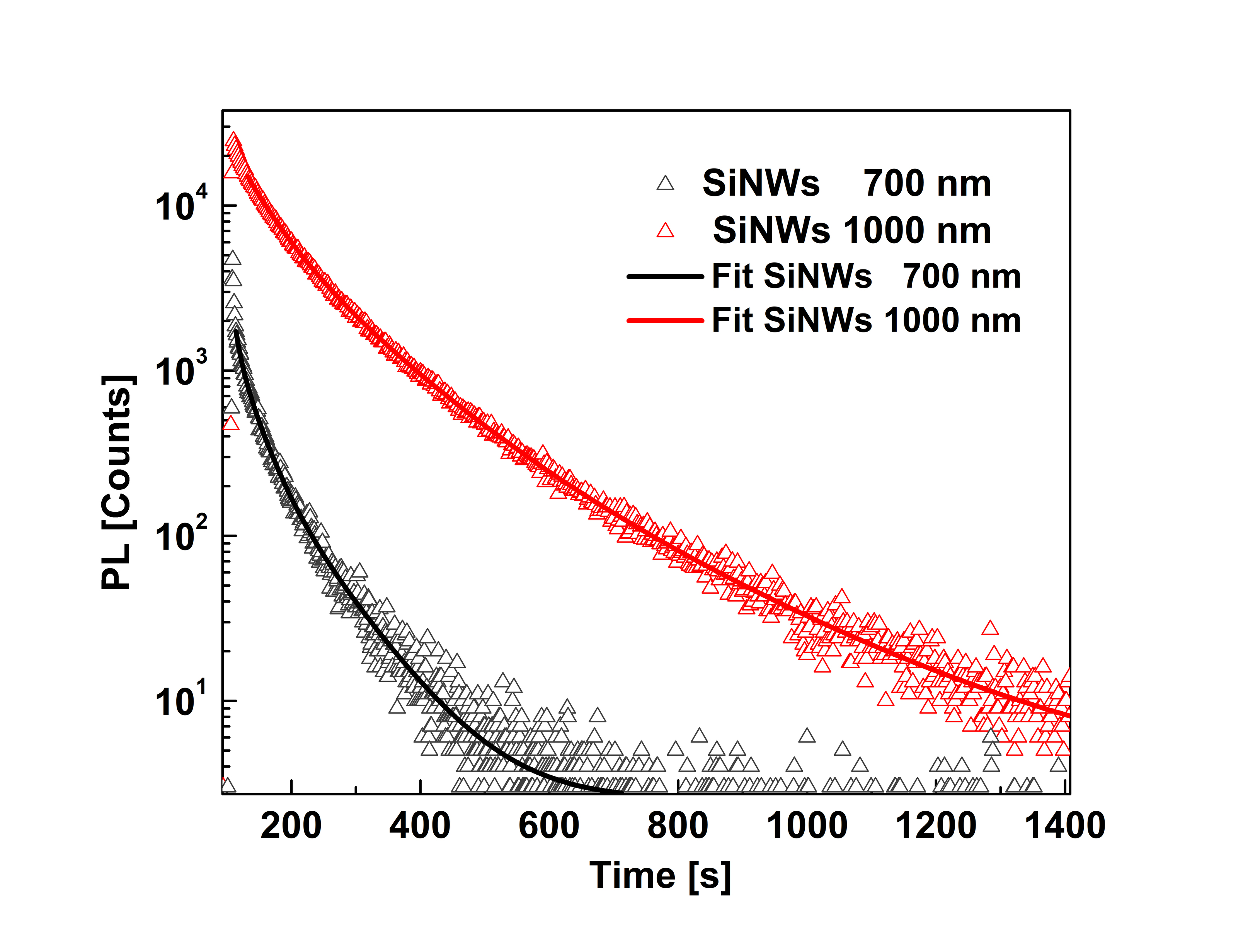}
    \caption{PL emission decay curves of the SiNWs arrays, with average lifetime 111 \textmu s  (1000 nm length) and 60 \textmu s (700 nm length), respectively.  } 
    \label{Figure 6}
\end{figure}

Fig.\ \ref{Figure 6} shows PL emission decay curves of the SiNWs arrays obtained by using the excitation wavelength 300 nm and the emission wavelength 770 nm. The monoexponential lifetime decreases sharply for shorter SiNWs, indicating a smaller contribution of the surface disorder. Long lifetime observed for the longer SiNWs should be mostly dictated by nonradiative processes involving surface defects, in agreement with the results of XRD-RSM maps presented in Figure 3. 
The average lifetime obtained by fitting the experimental values is 60 \textmu s for the short SiNWs and 111 \textmu s for long SiNWs. \textcolor{black}{By increasing the etching time, the number of both radiative and non-radiative centers increases, however their ratio remains relatively unchanged and that determines longer luminescence lifetime coupled with stronger radiative emission.}

\subsection{Electrical characteristics of Silicon nanowires}
\begin{figure}[H]
\vspace{-5mm}
  \begin{subfigure}{0.5\textwidth}
  \centering
  \includegraphics[height=11em,width=1\linewidth]{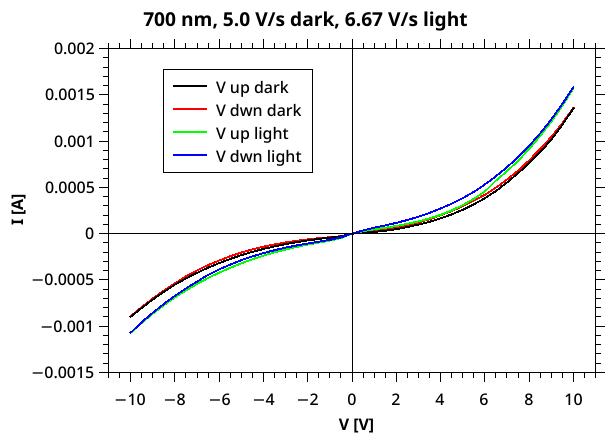} 
  \label{fig:sub-first} 
  \vspace{-11mm}
  \end{subfigure}
  \hfill
  \begin{subfigure}{0.5\textwidth}
  \centering
  \includegraphics[height=11em,width=1\linewidth]{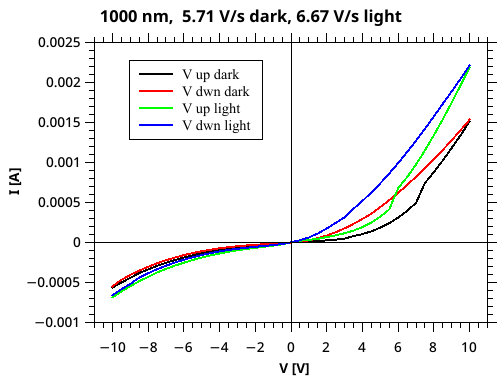}
  \label{fig:sub-second}
  \vspace{-11mm}
  \end{subfigure}
   (a)  \hspace{78 mm}   (f)  \vspace{2mm}
  \newline
  \begin{subfigure}{0.5\textwidth}
  \centering
  \includegraphics[height=11em,width=1\linewidth]{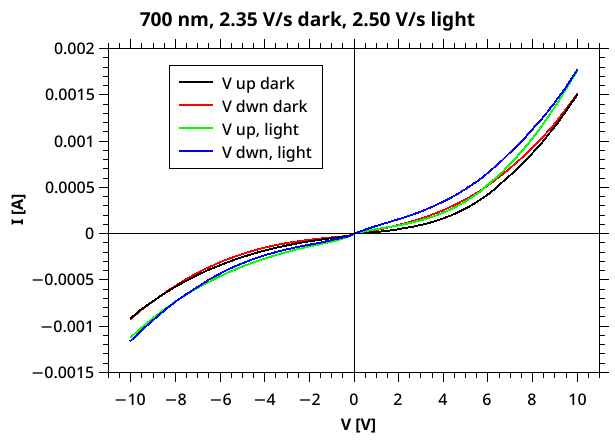}
  \label{fig:sub-third}
  \vspace{-11mm}
  \end{subfigure}
  \begin{subfigure}{0.5\textwidth}
  \centering
  \includegraphics[height=11em,width=1\linewidth]{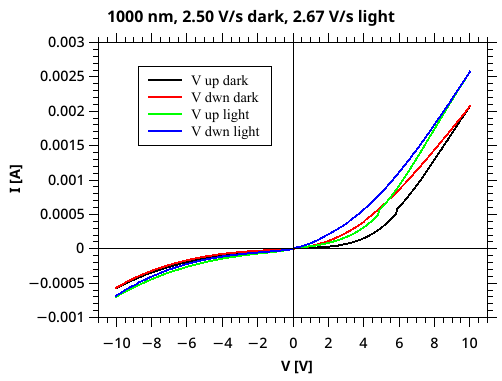}   
  \label{fig:sub-fourth}
  \vspace{-11mm}
  \end{subfigure}
  (b)  \hspace{78 mm}   (g)  \vspace{2mm}
  \newline
  \begin{subfigure}{0.5\textwidth}
  \centering
  \includegraphics[height=11em,width=1\linewidth]{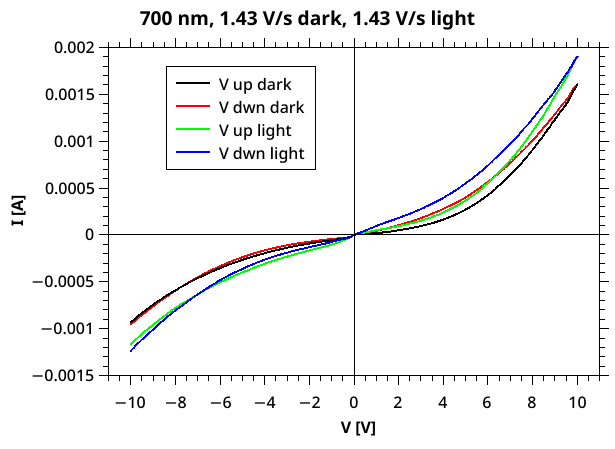}  
  \label{fig:sub-5}
  \vspace{-11mm}
 \end{subfigure}
 \begin{subfigure}{0.5\textwidth}
  \centering
  \includegraphics[height=11em,width=1\linewidth]{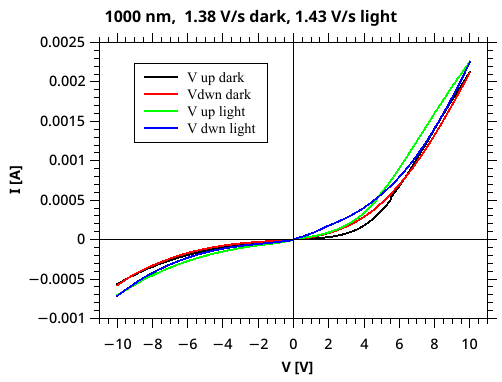}
  \label{fig:sub-6}
  \vspace{-11mm}
  \end{subfigure}
  (c)  \hspace{78 mm}   (h)  \vspace{2mm}
  \newline
  \begin{subfigure}{0.5\textwidth}
  \centering
  \includegraphics[height=11em,width=1\linewidth]{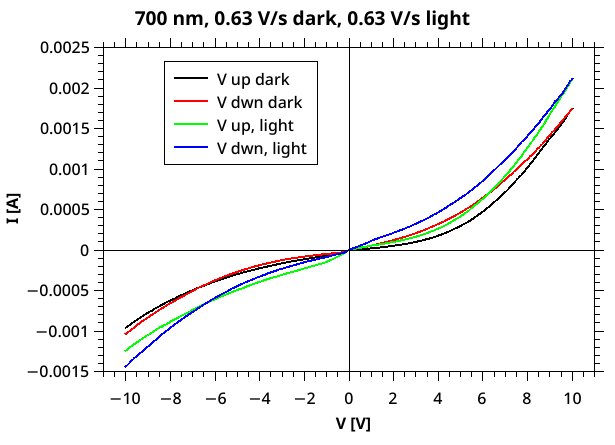} 
  \label{fig:sub-7}
  \vspace{-11mm}
  \end{subfigure}
  \begin{subfigure}{0.5\textwidth}
  \centering
  \includegraphics[height=11em,width=1\linewidth]{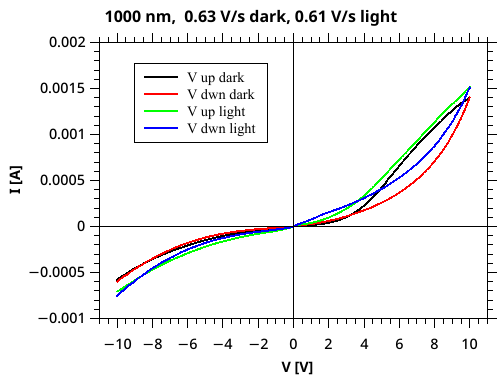}  
  \label{fig:sub-8}
  \vspace{-11mm}
  \end{subfigure}
   (d)  \hspace{78 mm}   (i)  \vspace{2mm}
  \newline
  \begin{subfigure}{0.5\textwidth}
  \centering
  \includegraphics[height=11em,width=1\linewidth]{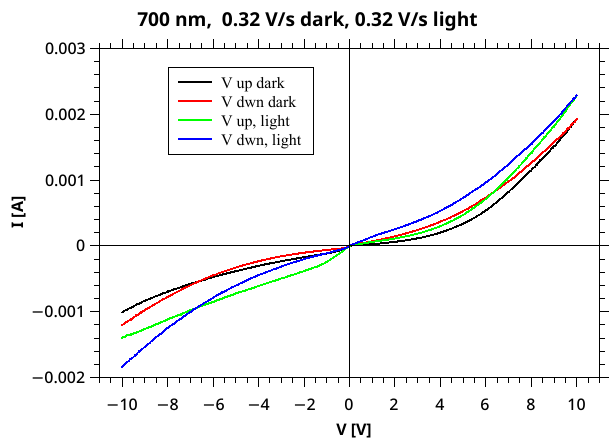}  
  \label{fig:sub-9}
  \vspace{-11mm}
  \end{subfigure}
  \begin{subfigure}{0.5\textwidth}
  \centering
  \includegraphics[height=11em,width=1\linewidth]{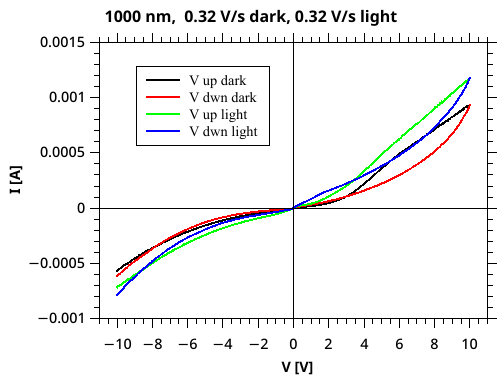}  
  \label{fig:sub-10}
  \vspace{-11mm}
  \end{subfigure}
   (e)  \hspace{78 mm}   (j)  \vspace{2mm}
  \caption{Current-voltage characteristics of Al/SiNWs/Al structures: (a)-(e) short SiNWs and (f)-(j) long SiNWs, measured at various voltage sweep rates. In dark and under illumination measurements.}
\label{fig7}
\end{figure}

\begin{figure}
    \centering
    \includegraphics[width=1\linewidth]{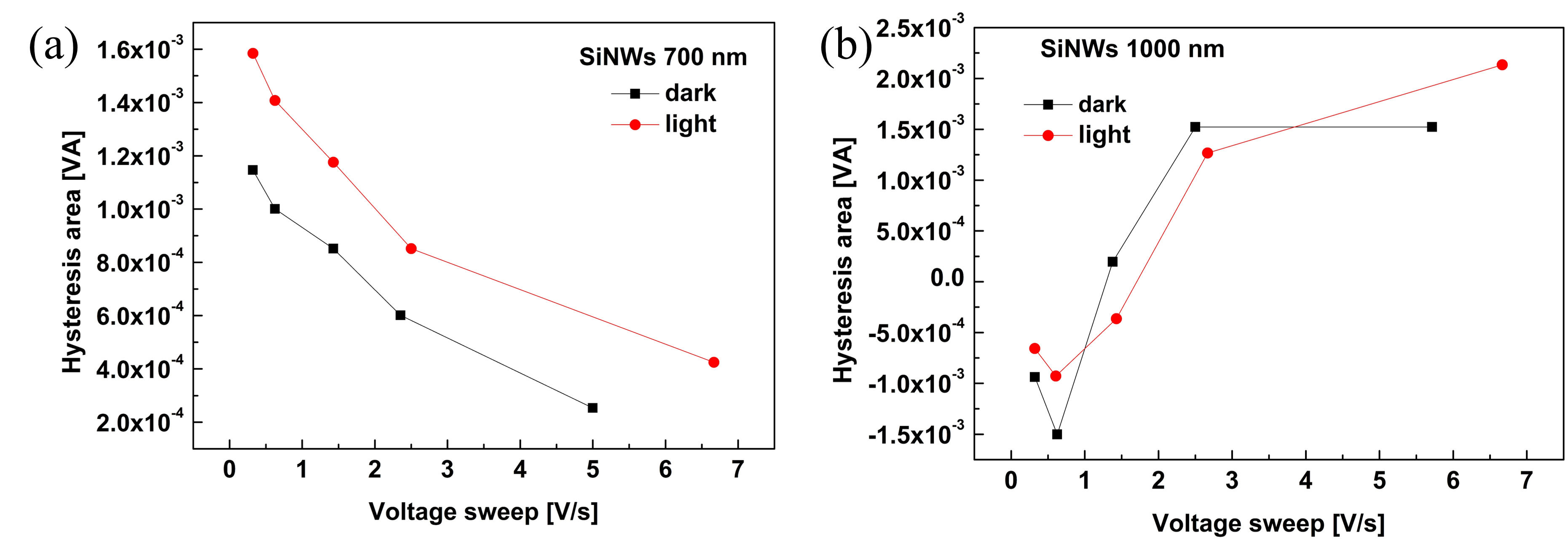}
    \caption{Hysteresis area vs voltage sweep rate for the device: (a) short SNWs and (b) long SiNWs.} 
    \label{Figure 8}
\end{figure}

The I-V curves of the Al/SiNWs/Al structures, measured at various voltage sweep rates in dark and under illumination, are presented in Figures 7 a-j. The plots are non-linear, characteristic to two diodes in antiparallel configuration, due to the Al-Si Schottky contacts. The current intensities under illumination are slightly higher compared to those measured in dark. The hysteresis observed in the I-V curves of both samples suggests that a process of trapping and de-trapping of minority (e) charge carriers, with different time constants, may take place. A dependence of the hysteresis area (in VA units) as a function of the voltage sweep rate may also be observed, see Fig.\ \ref{Figure 8}a,b. 

In the case of the SiNW$_{\mathrm {short}}$ sample the hysteresis area {\red in quadrant 1, defined as area of the ``down'' curve minus the area of the ``up'' curve}, shows a continuous decrease as the voltage sweep-rate increases, see Fig.\ \ref{Figure 8}a, but remains in the positive range of values. A different behavior may be observed in the case of SiNW$_{\mathrm {long}}$, as the hysteresis area takes negative values at small V rates and positive values for rates beyond ~1.38 V/s, see Fig.\ \ref{Figure 8}b. 
Also, the data reveal that the hysteresis area increases under illumination for short NWs, see Fig.\ \ref{Figure 8}a, but exhibits an interesting evolution at sweep rates below ~4 V/s in the case of long NWs, Fig.\ \ref{Figure 8}b. Additional results obtained on SiNW arrays with lengths larger than 1000 nm and exhibiting  different morphologies are shown in the Supplementary Material, Fig.\ \ref{SampleC}.
\textcolor{black}{where it has been observed that post-treated SiNWs with HF shows minimum hysteresis and confirm that surface defect-free nanowires can be prepared by MACE using an HF post-treatment. This result is in good agreement with Choi et al \citep{dawood2010interference}.
It is worth noting that the untreated bulk Si shows no hysteresis and confirms the larger defective surface area of SiNWs is associated with charge trapping and hysteresis effect. The corresponding plot is shown in Supplementary Material, Fig.\ \ref{bulk si}}
\textcolor{black}{As apparent, the extend of the hysteresis is not related with etching time in the range tested (2-10 min).}

\section{Modeling the I-V characteristics }

\begin{figure}
    \centering
    \includegraphics[width=0.7\linewidth]{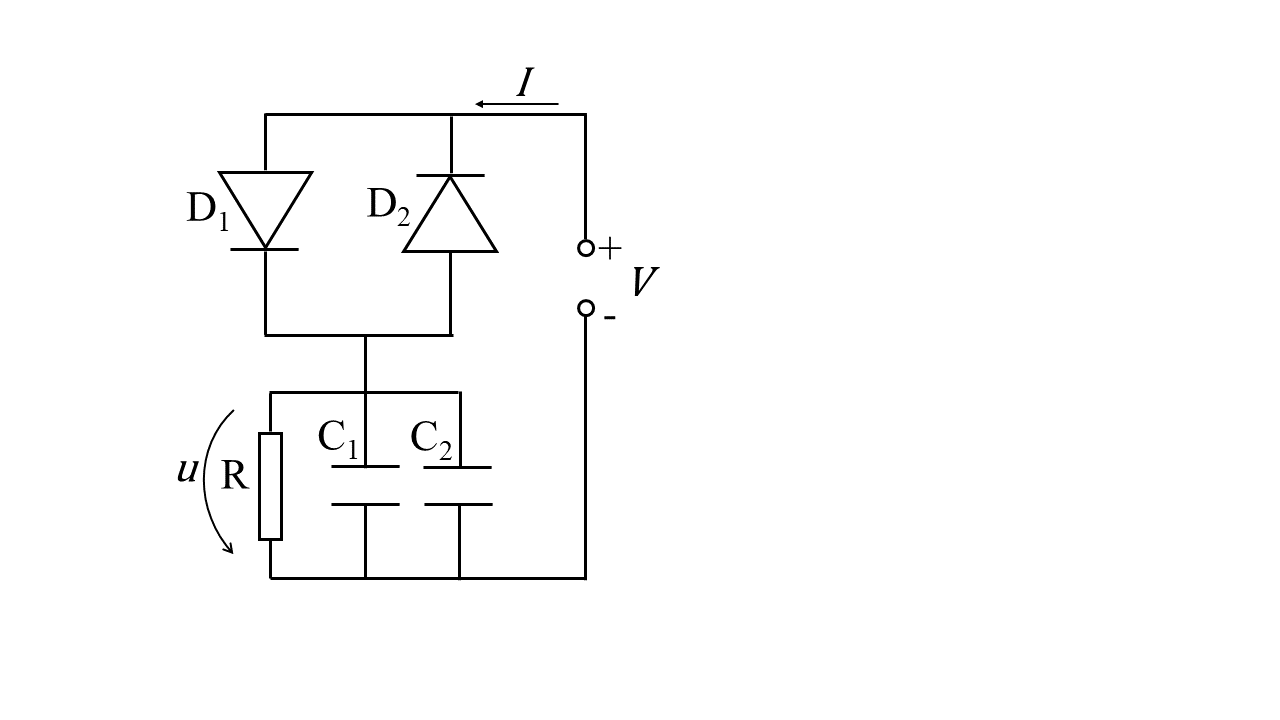}
    \caption{A simple circuit model for the I-V characteristic with hysteresis. The direction of the current depends on the polarity of the main voltage V(t).} 
    \label{circuit}
\end{figure}

We consider an equivalent electrical circuit of the Si/SiNWs/Al structures to model the \textcolor{black}{observed memristive behavior of measured I-V curves. The memristive effect consists in the dependence of the I-V curve, and consequently of the electrical resistance of the device, on the history of the applied voltage, i.\ e. increasing or decreasing \cite{Milano19}. I-V hysteresis loops for multiple cycles are shown in the Supplementary Material, Fig.\ \ref{Multicycle}.}
The I-V experimental data show exponential-like dependencies, and a difference between the current “up”, I$_{\mathrm {up}}$, i.e. when the voltage increases, and the current “down” I$_{\mathrm {down}}$, i.e. when the voltage decreases. A simple model to account for this behavior should combine a resistor, two diodes, and one or two capacitors, as illustrated in Figure\ \ref{circuit}.
The role of the diodes is to give the exponential-like current as a function of voltage, and the role of the $R-C$ block is to generate a voltage drop $u$ which controls the number of charge carriers passing through the circuit. The voltage $u$ corresponds to an electric field internal to the diodes, such that the currents through each diode $D_j,\ (j=1,2)$  can be written as:
\begin{equation}
    I_{D_{j}}(t)=I{_j}[e^\frac{q[V(t)-u(t)]}{n_{j}kT}-e^\frac{-qu(t)}{n_{j}kT}]
    \label{eq1} \ ,
\end{equation}
where $I_j$ and $n_j$ are diode parameters (saturation current and ideality factor, respectively), $q$ is the elementary electric charge, $k$ is Boltzmann’s constant, and $T$ the temperature.  We begin by considering only one capacitor, $C_1$, and ignore the second one, i.e. $C_2 = 0$.  The total current in the circuit is then:
\begin{equation}
  I = I_{D_{1}}+I_{D_{2}}=I_{R}+I_{C_{1}},
    \label{eq2}
\end{equation}
where $I_{R}=u/R$, and $I_{C_{1}}$ is discussed below.  

The order of magnitude of the charge associated with the capacitance $C_1$ needed to explain the experimental data can be inferred from the observed hysteresis effect. The total charge going through the circuit corresponds to the area of the current versus time, which can easily be evaluated since the voltage has a constant rate in time. For example, for the {\vio SiNW$_{\mathrm {short}}$} sample, at a voltage rate of 0.63 V/s in dark (Figure \ref{fig7}d), the hysteresis area between the positive voltages 4 V and 5 V, corresponding to a time interval of 1.6 s, is {\brown 0.15 mVA}, or  {\brown 0.24 mAs} or 0.24 mC of electric charge. This gives an estimated $C_1 \approx 0.24$  mF, which is obviously a very large value for such a small sample. A more realistic assumption is to assume a much smaller capacitance, and associate the capacitor with a trapping mechanism, which temporarily stores a relatively small amount of charge, $Q_1$, but contributes significantly to the voltage $u$, which in turn has a much larger effect on the current than the stored charge. {\red The capacitance is associated to a temporary polarization effect, likely due to interface states in the Schottky diodes \cite{Sze_Ch3} and/or surface states \cite{Sacchetto14,puppo2016surface}. } Therefore, we assume that the current controlled by the capacitor is
\begin{equation}
 I_{C_{1}}=b\frac{dQ_{1}}{dt}
\label{eq3}
\end{equation}
where b is a coefficient describing the amplification factor of the number of carriers controlled by the polarization effects associated with the charge Q$_1$. {\red Here Equation (\ref{eq3}) is a simplified version of Equation (5) of Reference \cite{puppo2016surface}, where both acceptor-like and donor-like traps are considered.}  

Next, we denote by {$\tau_1$} the time constant associated to the relaxation of this electric or trapping charge.  This relaxation process may depend on more complex phenomena, like ion displacement, diffusion, etc., which we cannot describe in detail.   Instead, we define the relaxation time {$\tau_1$} via the equation
\begin{equation}
   {\brown \frac{dQ_{1}}{dt}}=-\frac{Q_{1}(t)-C_{1}u(t)}{\tau_{1}}
    \label{eq4}
\end{equation}
which leads to an exponentially asymptotic charging or discharging with a time factor $e^{-t/\tau_1}$. {\red A similar assumption has been used to explain the hysteresis phenomenon in perovskite based solar cells \cite{nemnes2017dynamic}}.

\begin{figure}
    \centering
    \includegraphics[width=0.5\linewidth]{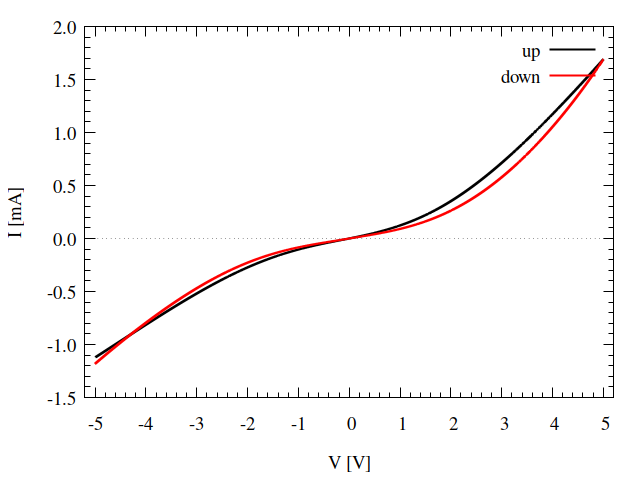}
    \caption{The I-V characteristic with a normal current $I_{C_1}$ .  When the voltage $V$ decreases (along the ``down'' curves) $I_{C_1}$ is oriented against the source, if it is released by a normal capacitor, and consequently the total current is smaller than it was when $V$ increased (the ``up'' curves).  }
    \label{fig10}
\end{figure}

In Fig.\ \ref{fig10} we show the calculated I-V characteristic using empirical parameters inspired by the experimental results, but also adjusted for the convenience of the numerical calculations: $I_1=0.04$ mA, $I_2=0.03$ mA, $n_1= -n_2=30$, $\tau_1=6$ s, $C_1=1.5$ nF, $b=10^6$,  $R=1600 \ \Omega$. The voltage is swept from -5 V to 5 V and back to -5 V in 50 seconds, i.e. with a rate of 0.4 V/s.  The current was calculated numerically using Equations (1)-(4), by discretizing the time in small steps, with initial conditions $Q_1=0$. (The current in the two diodes was obtained using the Lambert function, since the voltage $u(t)$ implicitly depends on the current.)
\\

\begin{figure}
    \centering
    \includegraphics[width=0.5\linewidth]{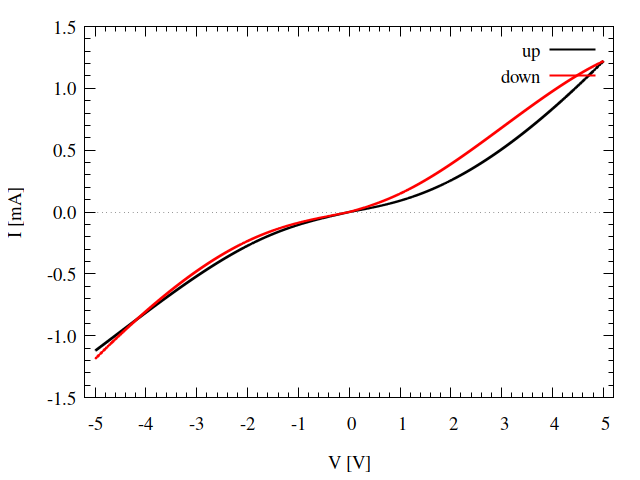}
    \caption{ The calculated I-V curves with the same parameters as in Figure 9, but negative {\brown coefficient} b.  In this case the current for positive voltage is smaller when the voltage decreases, as observed for the samples of 700 nm.}
    \label{fig11}
\end{figure}

The intersection of the ``up'' and ``down'' curves at negative voltages occurs because of the initial and final state of the capacitor (uncharged vs. charged).   However, one feature of the hysteresis loop shown in Fig.\ \ref{fig10} differs from the experiment. For positive voltage, the “up” curve is always above the down curve, i.e. opposite to the experimental data obtained for the sample A (700 nm).  The reason is that after the voltage reached the maximum and begins to decrease, the capacitor $C_1$ pushes current against the main current of the source, i.e. decreasing the current compared to the ``up'' values.  This effect does not depend on the magnitude of the coefficient $b$, but on its sign, which is positive.   To match the experimental data with this simple model we need to assume a {\em negative} sign of this coefficient $b$ for the ``up'' segment of the I-V characteristic, i.e. in that phase the trapping mechanism releases current with the same orientation as of the total current. The resulting I-V curves are shown in Fig.\ \ref{fig11}.

The I-V characteristic looks now qualitatively similar to the experimental data {\blue shown in Figures \ref{fig7}a-e}.  This similarity suggests a negative intrinsic polarization mechanism of the sample, during the measurements, with a relaxation time of the order of seconds. Such a situation can also be obtained in perovskite based solar cells, where the ``up'' and ``down'' currents in the hysteresis loops can be inverted, depending on the sign of the polarization of the cell \citep{tress2016inverted,nemnes2017dynamic}. {\vio It is also seen in Fig.~\ref{fig7} that in} presence of light the magnitude of the current increases, due to increased number of {\vio photogenerated} charge carriers.   

\begin{figure}
    \centering
    \includegraphics[width=0.5\linewidth]{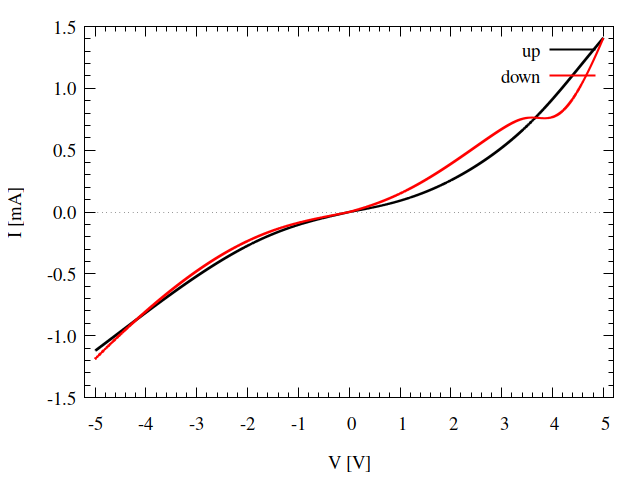}
    \caption{ The calculated I-V curves with the same parameters as in Figure 3, plus second capacitor {\brown $C_2$} = 0.6 nF which is activated gradually between voltages 0.5 - 1 V.  The current created by the second capacitor has always a positive coefficient b.  }
    \label{fig12}
\end{figure}

An additional feature may be observed in the I-V curves of {\vio the sample SiNW$_{\mathrm {long}}$}, where the hysteresis for positive voltage reverses with decreasing the voltage rate, see Figs.\ \ref{fig7}f-j.  An initial shoulder is visible at high rates on the ```up'' curve, below the ``down'' curve, which then moves above the ``down'' curves at lower voltage rates, below 1.38 V/s. {\red For lower voltage rates the hysteresis loop becomes twisted.} A possible interpretation is that in this situation another capacitor, $C_{2}$, is activated at a certain positive voltage, acting now in the regular manner, i.e. pumping current against the source, $I_{C_{2}}=b \ dQ_{2}/dt$ with the $b$ coefficient always positive. In Fig.\ \ref{fig12} we show the results with $C_2=0.6$ nF.  Such an example looks qualitatively similar with the data for the sample {\vio SiNW$_{\mathrm {long}}$} shown in Figs.\ \ref{fig7}h-j.  {\red Such a twisted hysteresis loop has been shown by Thissandier et al. \cite{THISSANDIER2012109}, for an array of disconnected SiNWs, when the voltage was increased above a certain threshold.}

We emphasize that the model used for explaining the I-V curves is primitive, and only qualitative.  The magnitude of the $b$ coefficient and of the capacitances have somehow a complementary character: we could increase one by decreasing the other one.  However, for a small $b$ value the capacitances would be unrealistic, and for this reason we believe their role is more like a trigger for activation of more charge carriers, typical for a small polarization field inside a Schottky or a p-n diode, or for a gate inside a transistor.  Since the NWs are formed in p-type Si, with resistivity  of the order of 1-10 $\Omega\cdot$cm, then shunt resistors should be included for a more realistic circuit model. Still, the development of a more complex equivalent circuit is beyond the scope of this study.

\subsection{Capacitance behavior under illumination}

\begin{figure}
    \centering
    \includegraphics[width=1\linewidth]{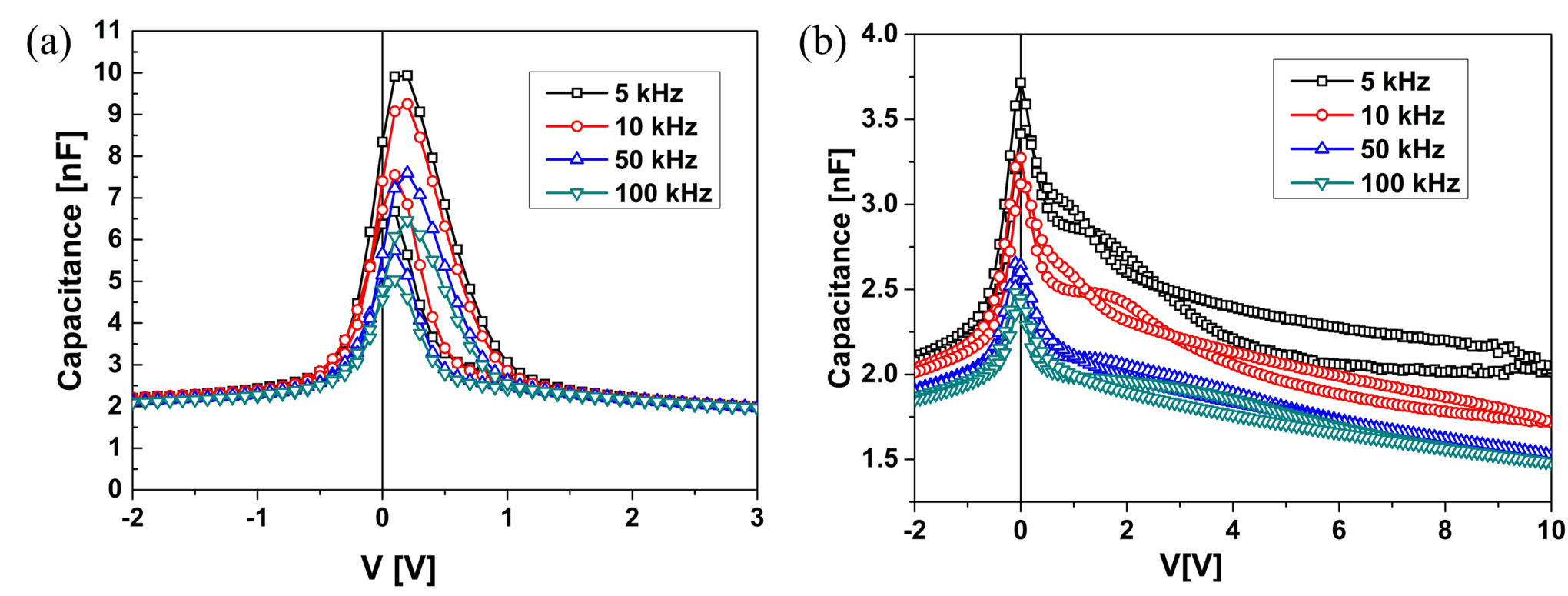}
    \caption{Capacitance-voltage characteristics of Al/SiNWs/Al structures at various frequencies: (a) SiNWs with 700 nm length and (b) SiNWs with 1000 nm lengths.}
    \label{fig14}
\end{figure}

\begin{figure}
    \centering
    \includegraphics[width=1\linewidth]{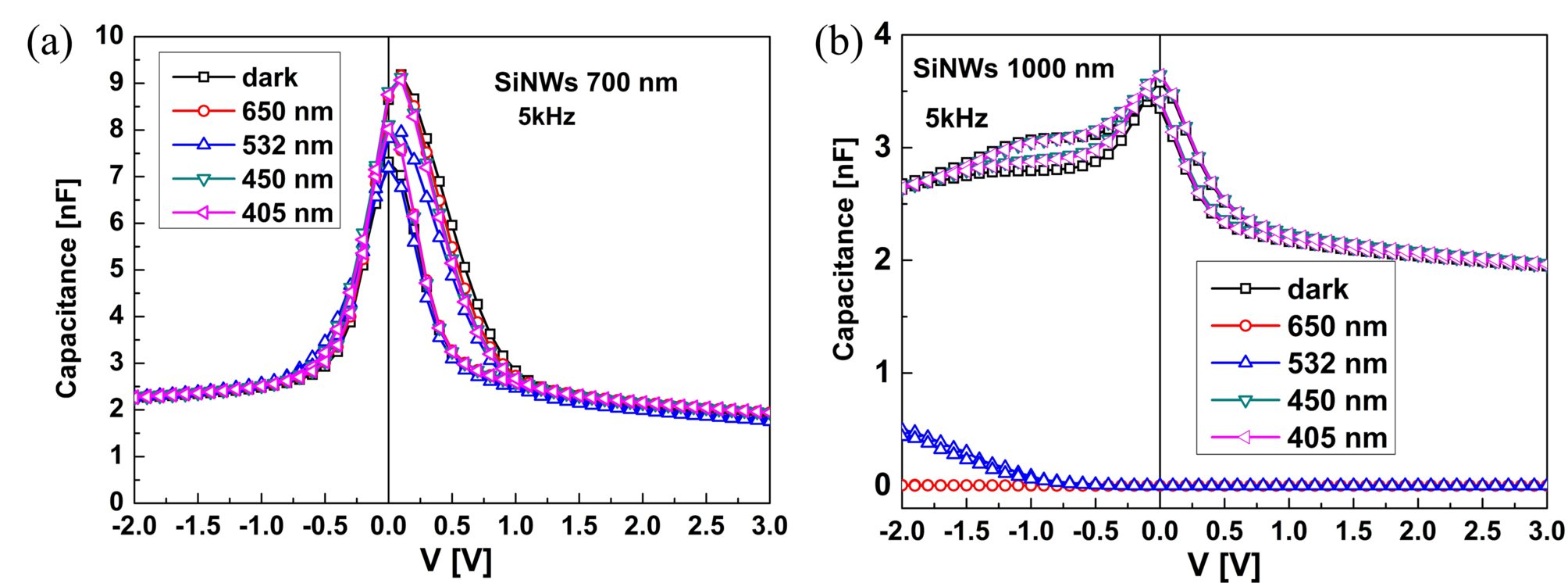}
    \caption{Capacitance-voltage characteristics at 5 kHz, of Al/SiNWs/Al structures with: a) {\vio short (700 nm)} NWs and b) {\vio long (1000 nm)} NWs, under illumination with various wavelengths.}
    \label{fig15}
\end{figure}

The C-V characteristics of the SiNW arrays measured at various frequencies are shown in Figure \ref{fig14}a,b. Both structures exhibit reduced capacitance by increasing the frequency in the range 5-100 kHz, with maximum value of 9.9 nF, Figure \ref{fig15}a, and 3.7 nF, Figure \ref{fig15}b, at 5 kHz. However, the SiNW arrays with long NWs exhibit asymmetric behavior in the region of positive voltages, 0 to +10V, where large and switching hysteresis loops appear under the forward and reverse polarization. The presence of a positive or negative sign of the current due to traps, or equivalently, negative or positive capacitance, as proposed in the equivalent circuit, could explain the observed behavior. 

Next we show the C-V characteristics measured under illumination with various wavelengths, in Fig.\ \ref{fig15}a,b. The capacitance of the structure with short NWs slightly increased when the structure is illuminated at 650 nm and 532 nm wavelength, Fig.\ \ref{fig15}a, while the structure with long NWs exhibits a significantly reduced capacitance under illumination at these wavelengths, Fig.\ \ref{fig15}b, likely due to the effect of  photogenerated carries trapped at the surface states, lowering the value of the capacitive reactance. Fig.\ \ref{fig15}b also shows the capacitance changing slightly under illumination with 450 nm or 405 nm wavelength. In this case the photogenerated carriers behave like free carriers and determine the increase of the current intensity, see I-V characteristics in Fig. \ref{fig7}. 

\begin{figure}
    \centering
    \includegraphics[width=1\linewidth]{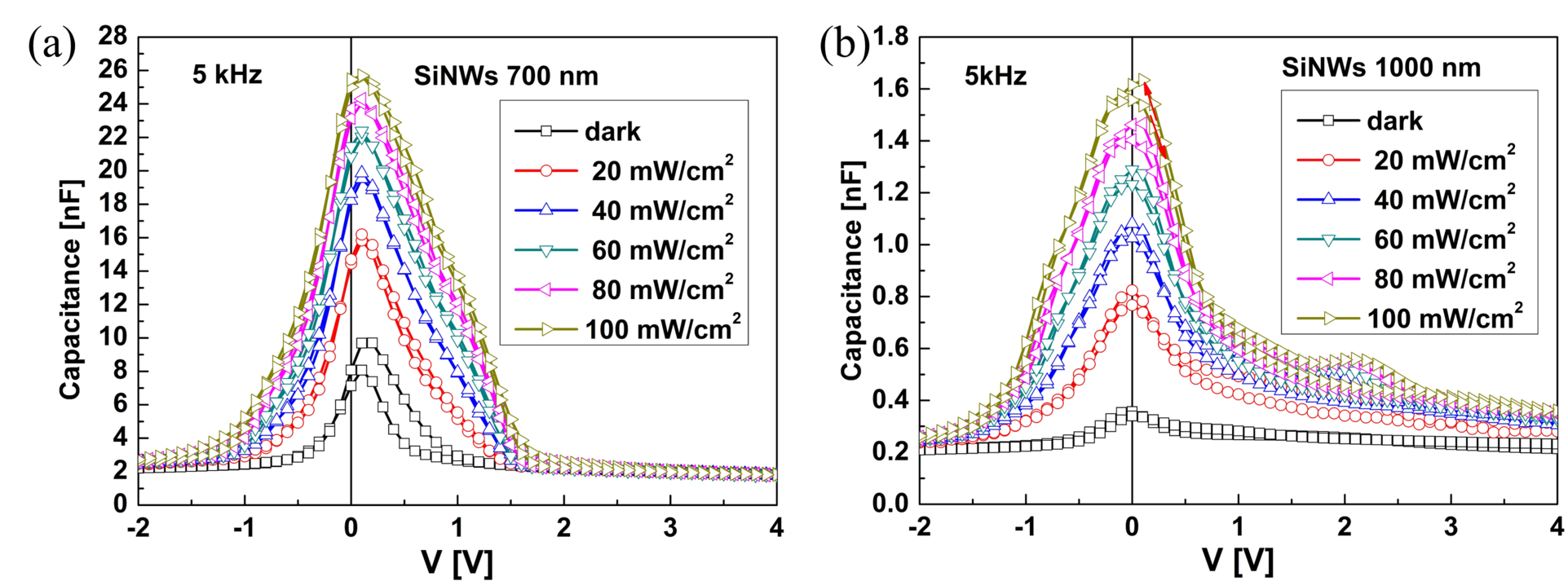}
    \caption{Capacitance-voltage characteristics at 5 kHz, of Al/SiNWs/Al structures with: a) short NWs and (b) long NWs, as function of the white light intensity.}
    \label{fig16}
\end{figure}

Fig.\ \ref{fig16} shows the C-V characteristics under illumination with white light, with various intensities. Both structures exhibit higher capacitance when the light intensity increases, reaching 26 nF for the structure with short NWs,  Fig.\ \ref{fig16}a, and 1.6 nF for SiNWs with long NWs, Fig.\ \ref{fig16}b,  under illumination with 100 mW/cm$^2$. All the C-V curves present hysteresis, noticeable under positive applied voltage, 0,+10V. The structure with long NWs shows also a supplementary capacitance peak at positive applied voltage, that increases with the intensity of light. Also, the occurrence voltage shifts towards higher values, from 1.82 V to 2.16 V with {\vio increased} light intensity.

\section{Discussion and conclusions}

The analysis of hysteresis exhibited by the I-V characteristics shown in Figs.~\ref{fig7}a-e, reveals that charge carriers are trapped on the surface states in sample {\vio SiNW$_{\mathrm {short}}$}, with PL average lifetime of 60 \textmu s, when the voltage sweep rate varies from 5 V/s to 0.32 V/s. The hysteresis area linearly increases by lowering the sweep rate, see Fig.~\ref{Figure 8}a, which suggests that the slow traps are involved in the process. The situation changes for {\vio the SiNW$_{\mathrm {long}}$} sample, with PL average lifetime of 111 \textmu s, where larger hysteresis areas appear at faster voltage sweep rates, Figs.~\ref{fig7}f-j. In fact, as the variation of the hysteresis area vs voltage sweep rate presented in Fig.~\ref{Figure 8}a,b suggests, at slow V rates, of 0.32 V/s and 0.63 V/s, the slow traps act in the both structures. Also, since faster voltage sweeps are required to activate the fast surface states, the evolution of the hysteresis area in Figure \ref{Figure 8}b indicates the presence of fast surface traps in the {\vio SiNW$_{\mathrm {long}}$} sample.  {\red The I-V hysteresis in SiNWs prepared by MACE has also been obtained in Ref.~\cite{Zaibi2022}, but the dependence on the voltage rate has not been reported. The presence of the traps in the SiNWs can also be inferred from the non-ideal diode characteristic of the Schottky contacts, observed in the I-V characteristic  \cite{Rouis2021}.}

The C-V response of this structure to excitation with various wavelengths suggests the presence of electron traps active at energy below 2.4 eV (observed in C-V responses at 1.9 eV and 2.33 eV irradiation), which are within the band gap of porous silicon \citep{rams1999cathodoluminescence,chen1994electrical,van1993photoelectron}. Although these band gap values are also within the (broad) emission band limits observed in Fig.~\ref{Figure 5}, they are different than the values derived from the maxima in the PL emission spectra, centered at 1.70 eV for {\vio SiNW$_{\mathrm {short}}$} and 1.61 eV for {\vio SiNW$_{\mathrm {long}}$}, suggesting that different trap centers are involved in the PL emission and photogeneration processes. 

In summary, the evolution of I-V hysteresis vs applied voltage rates was used to assess the effect of surface traps on transport properties of the pristine SiNWs arrays. The transition from an inverted to a direct hysteresis is demonstrated considering the effect of a {\red capacitance associated with the surface traps which controls the charge current through the structure.} The I-V and C-V characteristics measured under illumination with various wavelengths in the visible domain indicate that traps filling can be detected by tuning the photons energy. Using this method we assessed the energy range of the surface trap states energy of a SiNWs array.

\textcolor{black}{However, the characterization methods used in this work, including PL, I-V and CV, bring evidence on effects due to the presence of surface states, but hardly allow an accurate assignment of their energy to a specific defect or complex. First, taking into account that different techniques may give specific energy values to the same defect type, and second, the fact that we deal with a highly irregular surface which allows a variety of local environments and therefore prevents an accurate identification of the defect type, then a schematic representation of the associated energy levels may be misleading.}

\newpage 

\appendix*

\section{Supplementary Material}

\renewcommand{\thefigure}{A\arabic{figure}}
\setcounter{figure}{0}

\renewcommand{\theequation}{A\arabic{equation}}
\setcounter{equation}{0}

\begin{figure}[H]
    \centering
    \includegraphics[width=0.7\linewidth]{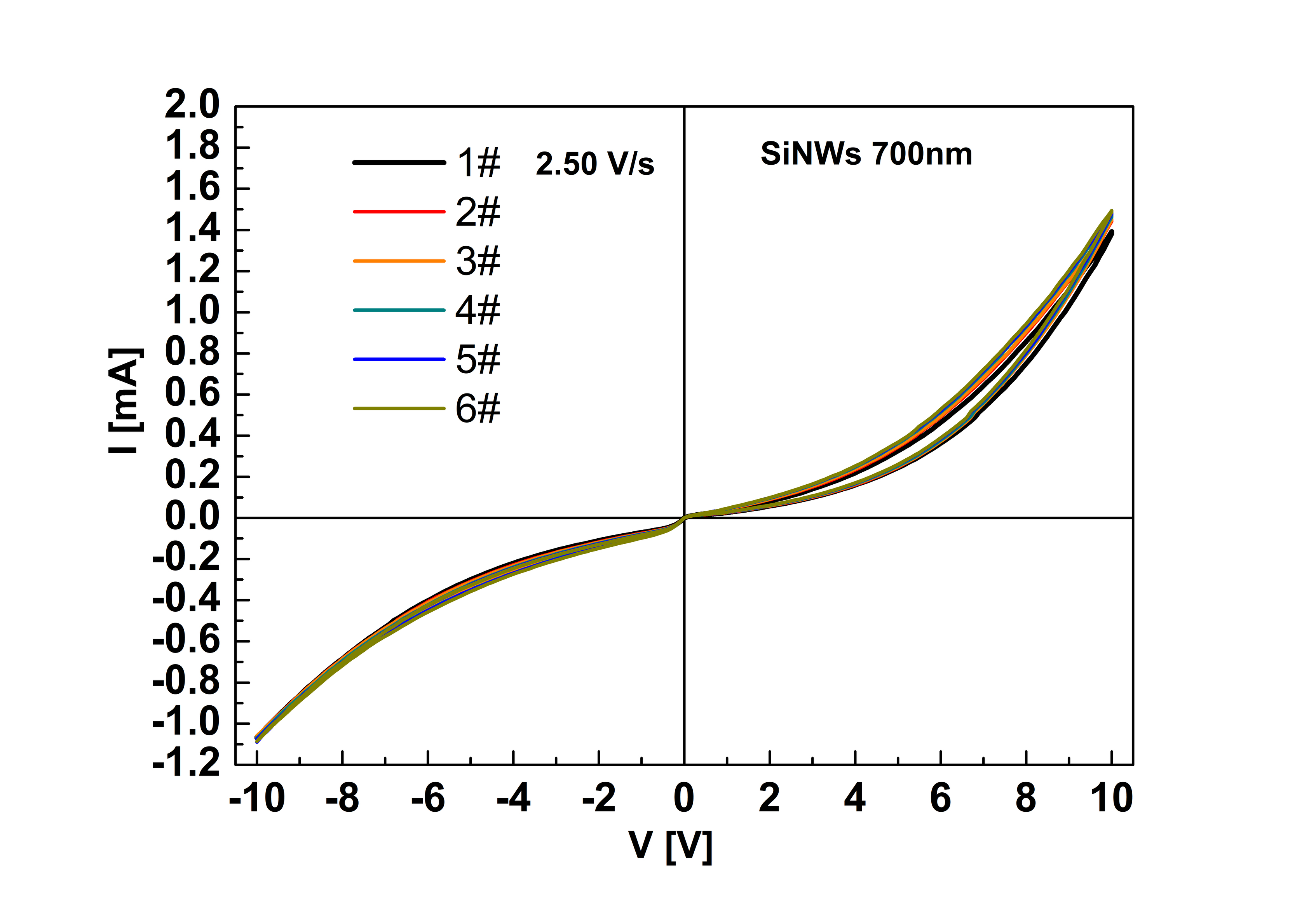}
    (a)
     \includegraphics[width=0.7\linewidth]{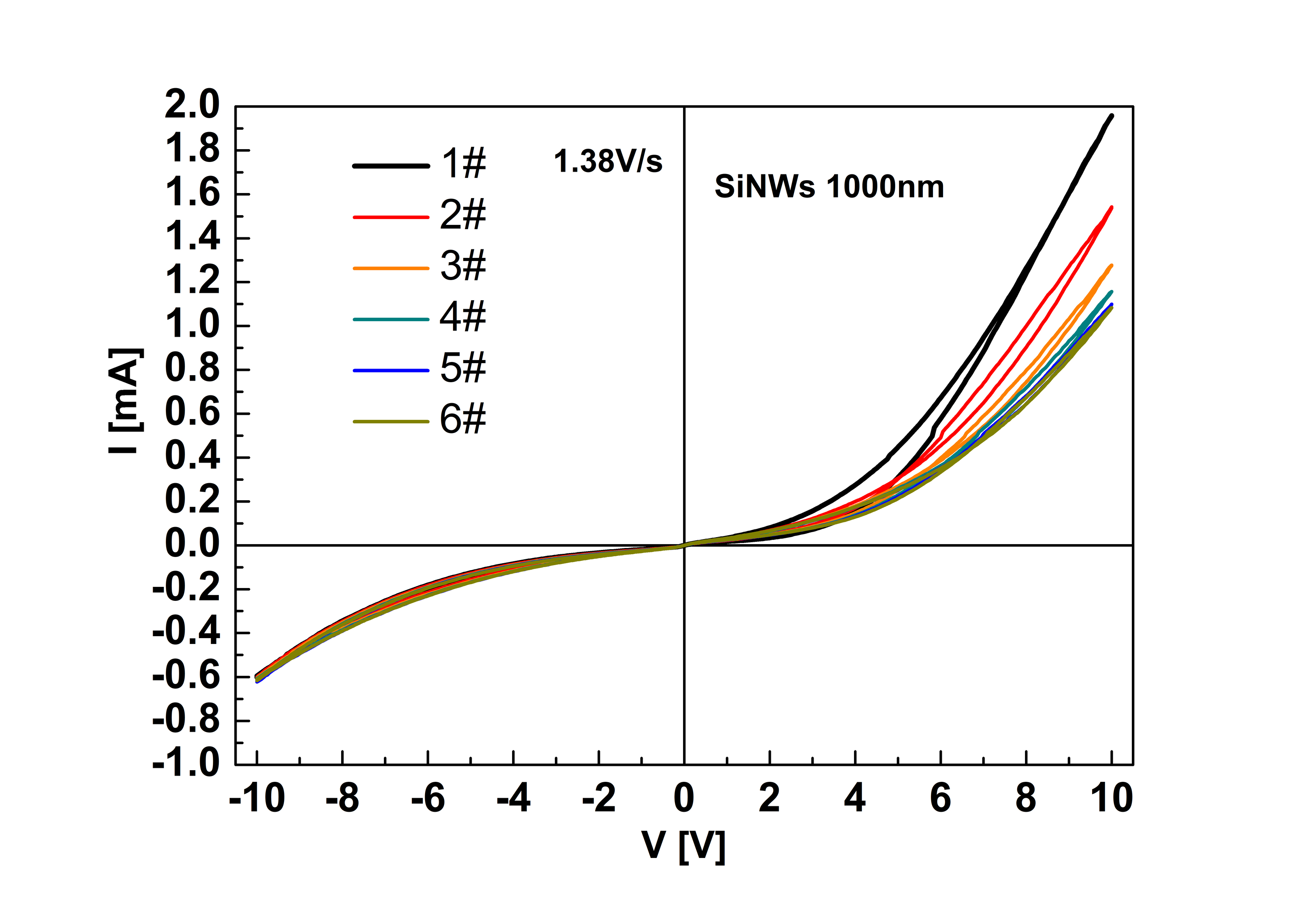}
    (b)
    \caption{Successive measurements of I-V characteristics for short (700 nm) (a), and long ( 1000 nm), (b),  SiNWs.}
    \label{Multicycle}
\end{figure}

 Successive measurements of I-V characteristics, displayed in Figures \ref{Multicycle} a,b, show that the hysteretic behavior is reproducible. A remanent charge localized on the surface state traps could determine a
slight shift of the hysteresis curves, as observed in the case of long SiNWs. The current intensity slightly increases, correlated with this initial trapping effect. Another possible explanation may
be the heating of the sample during the measurements.

\newpage

\begin{figure}[H]
    \centering
    \includegraphics[height=13em,width=0.6\linewidth]{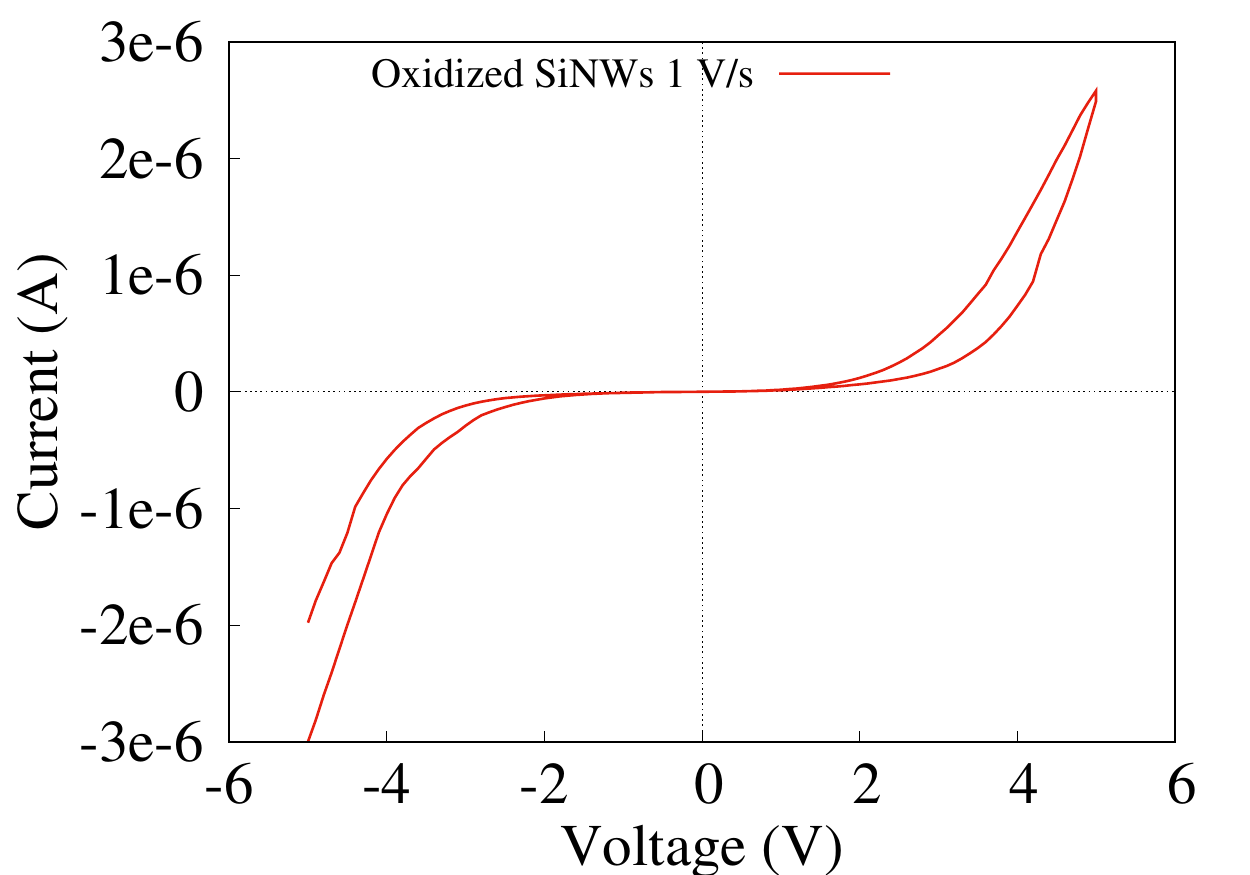}
    (a)
    \includegraphics[height=13em,width=0.6\linewidth]{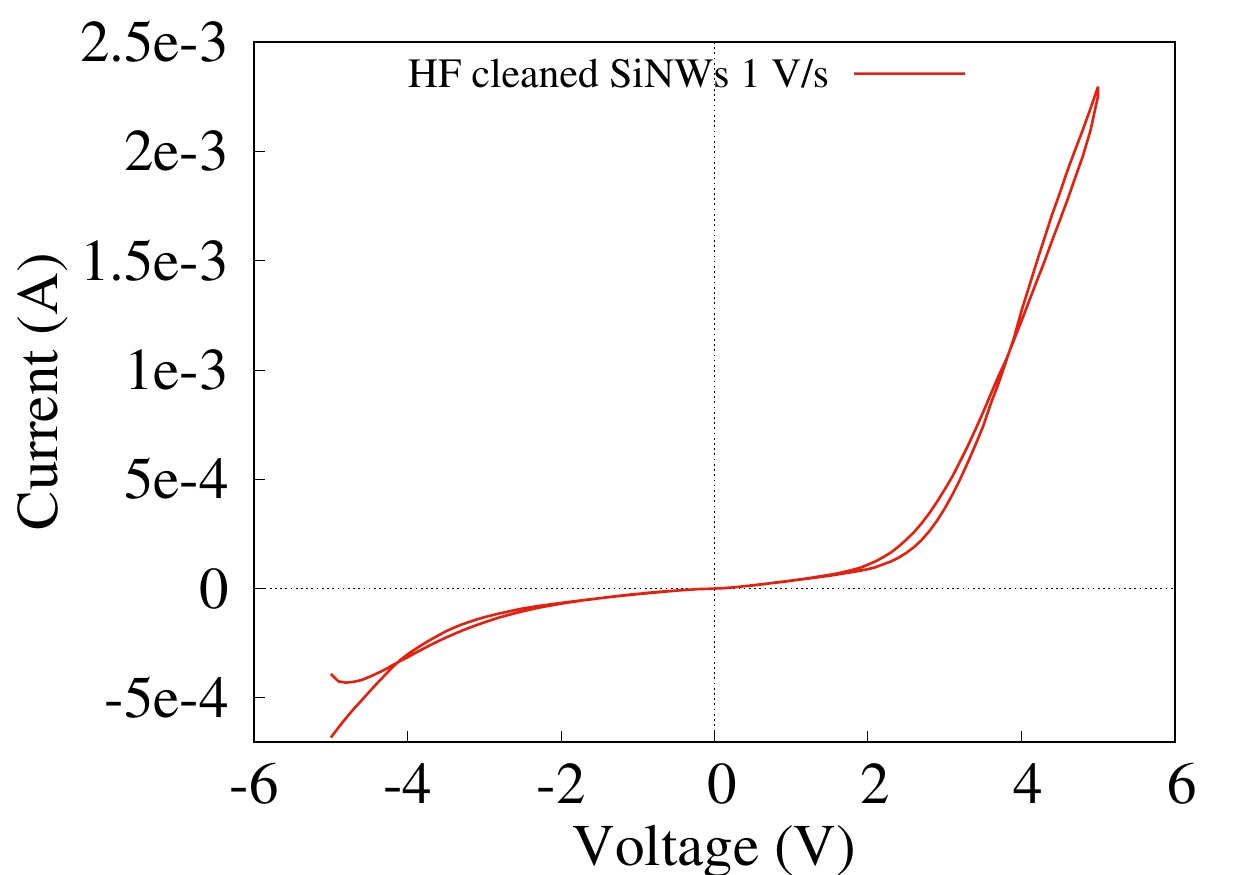}
    (b)
    \caption{I-V characteristics of oxidized (a), and HF cleaned (b), SiNWs. }
    \label{SampleC}
\end{figure}

I-V curves measured on a third series of samples with 2 \textmu m NW length are shown in Figures \ref{SampleC} a,b. The two SiNW arrays were processed with the same etching time, 10 minutes. Sample (a) oxidized at
room temperature prior to contacts deposition. Sample (b) was treated in HF prior to contacts deposition, in order to remove any oxide layer from the surface of SiNWs. Sample (b) shows a smaller hysteresis area than that of sample (a), also twisted hysteresis in I-V characteristics, with 3 orders of magnitude smaller resistance. These results relate the hysteresis to the presence of trapping states at the  
Si-SiO$_{2}$ interfaces and underline their contribution to the conduction mechanism.

\newpage

\begin{figure}[H]
    \centering
    \includegraphics[height=13em,width=0.6\linewidth]{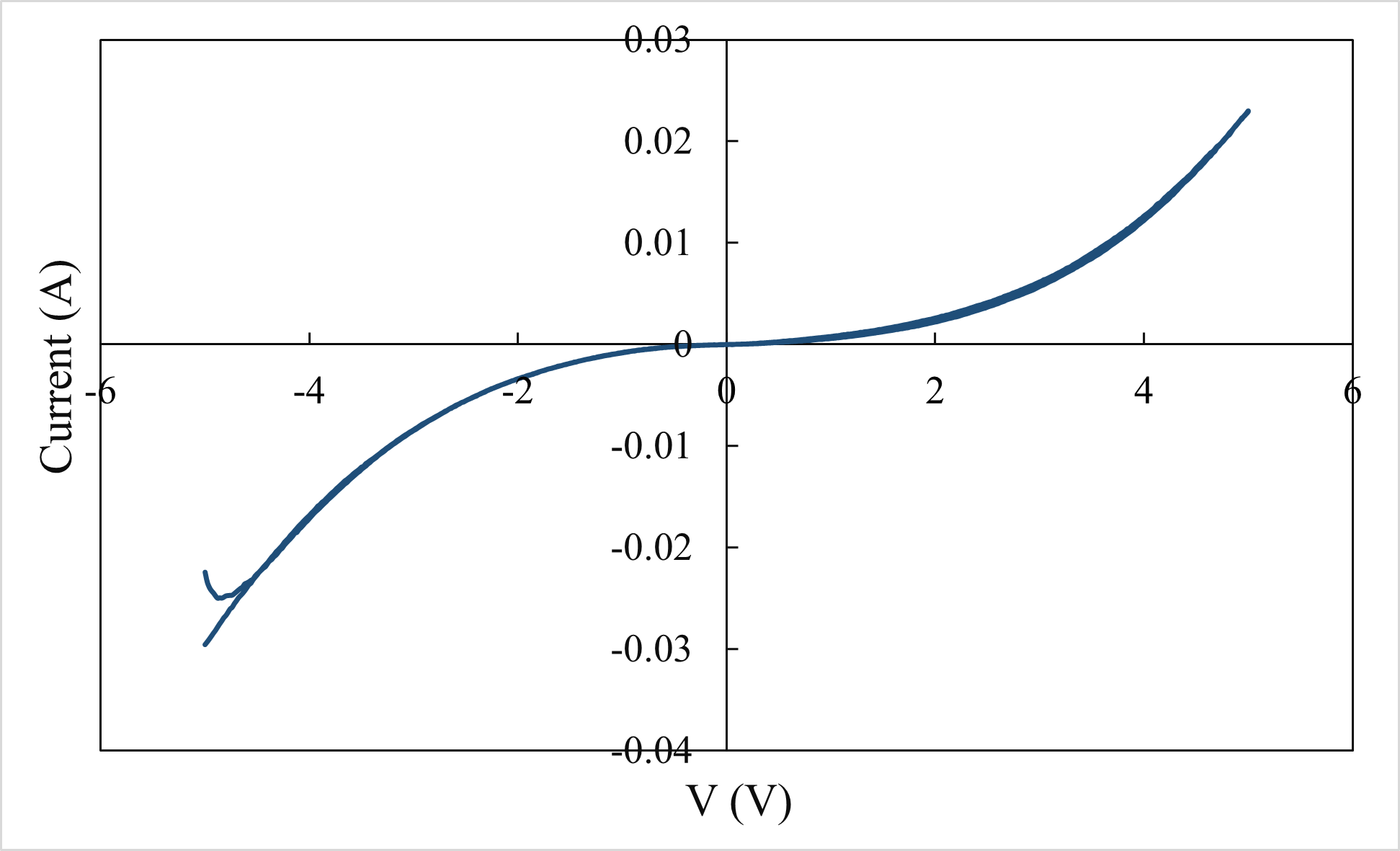}
    \caption{I-V characteristics of untreated bulk Si.}
    \label{bulk si}
\end{figure}

The I-V curves of bulk silicon, plotted in the Figures \ref{bulk si}, show no hysteresis. The measurement has been done as a reference and supports that the hysteresis observed in the corresponding plots of SiNW arrays are related to the much larger and more defective surface area of SiNWs.

\section*{Acknowledgments}
This work is partially funded by Reykjavik University Ph.D. fund no. 220006 and the Icelandic Research Fund Grant no. 218029-051. RP acknowledges support from the Core program 14N/2019 MICRO-NANO-SIS PLUS.

\section*{Credit Author Statement}
Author Contributions: Conceptualization, RP, HGS, AM; methodology, investigation, EF, MTS, HÖÁ (MACE processes, electrical measurements), CR (x-ray diffraction), IM (PL), GC (SEM), RP, EF, HÖÁ (electrical characterization); modeling, AM, GAN.; formal analysis, RP, AM; data curation, EF, HGS; writing and original draft preparation, RP, NP, AM; writing, review and editing, RP, NP, AM, HGS, SI; visualization, EF, CR, IM, RP; funding acquisition, HGS, SI, and RP. 

\section*{Declaration of Competing Interest}
All authors have read and agreed to the published version of the manuscript. The authors declare no conflict of interest.

\bibliography{Bibliography2}

\begin{thebibliography}{41}%
\makeatletter
\providecommand \@ifxundefined [1]{%
 \@ifx{#1\undefined}
}%
\providecommand \@ifnum [1]{%
 \ifnum #1\expandafter \@firstoftwo
 \else \expandafter \@secondoftwo
 \fi
}%
\providecommand \@ifx [1]{%
 \ifx #1\expandafter \@firstoftwo
 \else \expandafter \@secondoftwo
 \fi
}%
\providecommand \natexlab [1]{#1}%
\providecommand \enquote  [1]{``#1''}%
\providecommand \bibnamefont  [1]{#1}%
\providecommand \bibfnamefont [1]{#1}%
\providecommand \citenamefont [1]{#1}%
\providecommand \href@noop [0]{\@secondoftwo}%
\providecommand \href [0]{\begingroup \@sanitize@url \@href}%
\providecommand \@href[1]{\@@startlink{#1}\@@href}%
\providecommand \@@href[1]{\endgroup#1\@@endlink}%
\providecommand \@sanitize@url [0]{\catcode `\\12\catcode `\$12\catcode
  `\&12\catcode `\#12\catcode `\^12\catcode `\_12\catcode `\%12\relax}%
\providecommand \@@startlink[1]{}%
\providecommand \@@endlink[0]{}%
\providecommand \url  [0]{\begingroup\@sanitize@url \@url }%
\providecommand \@url [1]{\endgroup\@href {#1}{\urlprefix }}%
\providecommand \urlprefix  [0]{URL }%
\providecommand \Eprint [0]{\href }%
\providecommand \doibase [0]{http://dx.doi.org/}%
\providecommand \selectlanguage [0]{\@gobble}%
\providecommand \bibinfo  [0]{\@secondoftwo}%
\providecommand \bibfield  [0]{\@secondoftwo}%
\providecommand \translation [1]{[#1]}%
\providecommand \BibitemOpen [0]{}%
\providecommand \bibitemStop [0]{}%
\providecommand \bibitemNoStop [0]{.\EOS\space}%
\providecommand \EOS [0]{\spacefactor3000\relax}%
\providecommand \BibitemShut  [1]{\csname bibitem#1\endcsname}%
\let\auto@bib@innerbib\@empty
\bibitem [{\citenamefont {Chien}\ \emph {et~al.}(2020)\citenamefont {Chien},
  \citenamefont {Wei},\ and\ \citenamefont {Chen}}]{chien2020high}%
  \BibitemOpen
  \bibfield  {author} {\bibinfo {author} {\bibfnamefont {P.-J.}\ \bibnamefont
  {Chien}}, \bibinfo {author} {\bibfnamefont {T.-C.}\ \bibnamefont {Wei}}, \
  and\ \bibinfo {author} {\bibfnamefont {C.-Y.}\ \bibnamefont {Chen}},\
  }\href@noop {} {\bibfield  {journal} {\bibinfo  {journal} {Nanoscale research
  letters}\ }\textbf {\bibinfo {volume} {15}},\ \bibinfo {pages} {1} (\bibinfo
  {year} {2020})}\BibitemShut {NoStop}%
\bibitem [{\citenamefont {Han}\ \emph {et~al.}(2014)\citenamefont {Han},
  \citenamefont {Huang},\ and\ \citenamefont {Lee}}]{han2014metal}%
  \BibitemOpen
  \bibfield  {author} {\bibinfo {author} {\bibfnamefont {H.}~\bibnamefont
  {Han}}, \bibinfo {author} {\bibfnamefont {Z.}~\bibnamefont {Huang}}, \ and\
  \bibinfo {author} {\bibfnamefont {W.}~\bibnamefont {Lee}},\ }\href@noop {}
  {\bibfield  {journal} {\bibinfo  {journal} {Nano today}\ }\textbf {\bibinfo
  {volume} {9}},\ \bibinfo {pages} {271} (\bibinfo {year} {2014})}\BibitemShut
  {NoStop}%
\bibitem [{\citenamefont {Azeredo}\ \emph {et~al.}(2013)\citenamefont
  {Azeredo}, \citenamefont {Sadhu}, \citenamefont {Ma}, \citenamefont {Jacobs},
  \citenamefont {Kim}, \citenamefont {Lee}, \citenamefont {Eraker},
  \citenamefont {Li}, \citenamefont {Sinha}, \citenamefont {Fang} \emph
  {et~al.}}]{azeredo2013silicon}%
  \BibitemOpen
  \bibfield  {author} {\bibinfo {author} {\bibfnamefont {B.}~\bibnamefont
  {Azeredo}}, \bibinfo {author} {\bibfnamefont {J.}~\bibnamefont {Sadhu}},
  \bibinfo {author} {\bibfnamefont {J.}~\bibnamefont {Ma}}, \bibinfo {author}
  {\bibfnamefont {K.}~\bibnamefont {Jacobs}}, \bibinfo {author} {\bibfnamefont
  {J.}~\bibnamefont {Kim}}, \bibinfo {author} {\bibfnamefont {K.}~\bibnamefont
  {Lee}}, \bibinfo {author} {\bibfnamefont {J.}~\bibnamefont {Eraker}},
  \bibinfo {author} {\bibfnamefont {X.}~\bibnamefont {Li}}, \bibinfo {author}
  {\bibfnamefont {S.}~\bibnamefont {Sinha}}, \bibinfo {author} {\bibfnamefont
  {N.}~\bibnamefont {Fang}},  \emph {et~al.},\ }\href@noop {} {\bibfield
  {journal} {\bibinfo  {journal} {Nanotechnology}\ }\textbf {\bibinfo {volume}
  {24}},\ \bibinfo {pages} {225305} (\bibinfo {year} {2013})}\BibitemShut
  {NoStop}%
\bibitem [{\citenamefont {Baraban}\ \emph {et~al.}(2019)\citenamefont
  {Baraban}, \citenamefont {Ibarlucea}, \citenamefont {Baek},\ and\
  \citenamefont {Cuniberti}}]{baraban2019hybrid}%
  \BibitemOpen
  \bibfield  {author} {\bibinfo {author} {\bibfnamefont {L.}~\bibnamefont
  {Baraban}}, \bibinfo {author} {\bibfnamefont {B.}~\bibnamefont {Ibarlucea}},
  \bibinfo {author} {\bibfnamefont {E.}~\bibnamefont {Baek}}, \ and\ \bibinfo
  {author} {\bibfnamefont {G.}~\bibnamefont {Cuniberti}},\ }\href@noop {}
  {\bibfield  {journal} {\bibinfo  {journal} {Advanced Science}\ }\textbf
  {\bibinfo {volume} {6}},\ \bibinfo {pages} {1900522} (\bibinfo {year}
  {2019})}\BibitemShut {NoStop}%
\bibitem [{\citenamefont {Guan}\ \emph {et~al.}(2021)\citenamefont {Guan},
  \citenamefont {Cao},\ and\ \citenamefont {Li}}]{guan2021single}%
  \BibitemOpen
  \bibfield  {author} {\bibinfo {author} {\bibfnamefont {Y.}~\bibnamefont
  {Guan}}, \bibinfo {author} {\bibfnamefont {G.}~\bibnamefont {Cao}}, \ and\
  \bibinfo {author} {\bibfnamefont {X.}~\bibnamefont {Li}},\ }\href@noop {}
  {\bibfield  {journal} {\bibinfo  {journal} {Applied Physics Letters}\
  }\textbf {\bibinfo {volume} {118}},\ \bibinfo {pages} {153904} (\bibinfo
  {year} {2021})}\BibitemShut {NoStop}%
\bibitem [{\citenamefont {Za{\"i}bi}\ \emph {et~al.}(2022)\citenamefont
  {Za{\"i}bi}, \citenamefont {Slama}, \citenamefont {Beshchasna}, \citenamefont
  {Opitz}, \citenamefont {Mkandawire},\ and\ \citenamefont
  {Chtourou}}]{Zaibi2022}%
  \BibitemOpen
  \bibfield  {author} {\bibinfo {author} {\bibfnamefont {F.}~\bibnamefont
  {Za{\"i}bi}}, \bibinfo {author} {\bibfnamefont {I.}~\bibnamefont {Slama}},
  \bibinfo {author} {\bibfnamefont {N.}~\bibnamefont {Beshchasna}}, \bibinfo
  {author} {\bibfnamefont {J.}~\bibnamefont {Opitz}}, \bibinfo {author}
  {\bibfnamefont {M.}~\bibnamefont {Mkandawire}}, \ and\ \bibinfo {author}
  {\bibfnamefont {R.}~\bibnamefont {Chtourou}},\ }\href@noop {} {\bibfield
  {journal} {\bibinfo  {journal} {Journal of Applied Electrochemistry}\
  }\textbf {\bibinfo {volume} {52}},\ \bibinfo {pages} {273} (\bibinfo {year}
  {2022})}\BibitemShut {NoStop}%
\bibitem [{\citenamefont {Moh}\ \emph {et~al.}(2013)\citenamefont {Moh},
  \citenamefont {Nie}, \citenamefont {Pandraud}, \citenamefont {de~Smet},
  \citenamefont {Sudh{\"o}lter}, \citenamefont {Huang},\ and\ \citenamefont
  {Sarro}}]{moh2013effect}%
  \BibitemOpen
  \bibfield  {author} {\bibinfo {author} {\bibfnamefont {T.}~\bibnamefont
  {Moh}}, \bibinfo {author} {\bibfnamefont {M.}~\bibnamefont {Nie}}, \bibinfo
  {author} {\bibfnamefont {G.}~\bibnamefont {Pandraud}}, \bibinfo {author}
  {\bibfnamefont {L.}~\bibnamefont {de~Smet}}, \bibinfo {author} {\bibfnamefont
  {E.}~\bibnamefont {Sudh{\"o}lter}}, \bibinfo {author} {\bibfnamefont
  {Q.}~\bibnamefont {Huang}}, \ and\ \bibinfo {author} {\bibfnamefont
  {P.}~\bibnamefont {Sarro}},\ }\href@noop {} {\bibfield  {journal} {\bibinfo
  {journal} {Electronics letters}\ }\textbf {\bibinfo {volume} {49}},\ \bibinfo
  {pages} {782} (\bibinfo {year} {2013})}\BibitemShut {NoStop}%
\bibitem [{\citenamefont {Dan}\ \emph {et~al.}(2011)\citenamefont {Dan},
  \citenamefont {Seo}, \citenamefont {Takei}, \citenamefont {Meza},
  \citenamefont {Javey},\ and\ \citenamefont {Crozier}}]{dan2011dramatic}%
  \BibitemOpen
  \bibfield  {author} {\bibinfo {author} {\bibfnamefont {Y.}~\bibnamefont
  {Dan}}, \bibinfo {author} {\bibfnamefont {K.}~\bibnamefont {Seo}}, \bibinfo
  {author} {\bibfnamefont {K.}~\bibnamefont {Takei}}, \bibinfo {author}
  {\bibfnamefont {J.~H.}\ \bibnamefont {Meza}}, \bibinfo {author}
  {\bibfnamefont {A.}~\bibnamefont {Javey}}, \ and\ \bibinfo {author}
  {\bibfnamefont {K.~B.}\ \bibnamefont {Crozier}},\ }\href@noop {} {\bibfield
  {journal} {\bibinfo  {journal} {Nano letters}\ }\textbf {\bibinfo {volume}
  {11}},\ \bibinfo {pages} {2527} (\bibinfo {year} {2011})}\BibitemShut
  {NoStop}%
\bibitem [{\citenamefont {Milano}\ \emph {et~al.}(2019)\citenamefont {Milano},
  \citenamefont {Porro}, \citenamefont {Valov},\ and\ \citenamefont
  {Ricciardi}}]{Milano19}%
  \BibitemOpen
  \bibfield  {author} {\bibinfo {author} {\bibfnamefont {G.}~\bibnamefont
  {Milano}}, \bibinfo {author} {\bibfnamefont {S.}~\bibnamefont {Porro}},
  \bibinfo {author} {\bibfnamefont {I.}~\bibnamefont {Valov}}, \ and\ \bibinfo
  {author} {\bibfnamefont {C.}~\bibnamefont {Ricciardi}},\ }\href@noop {}
  {\bibfield  {journal} {\bibinfo  {journal} {Advanced Electronic Materials}\
  }\textbf {\bibinfo {volume} {5}},\ \bibinfo {pages} {1800909} (\bibinfo
  {year} {2019})}\BibitemShut {NoStop}%
\bibitem [{\citenamefont {Carrara}\ \emph {et~al.}(2012)\citenamefont
  {Carrara}, \citenamefont {Sacchetto}, \citenamefont {Doucey}, \citenamefont
  {Baj-Rossi}, \citenamefont {De~Micheli},\ and\ \citenamefont
  {Leblebici}}]{carrara2012memristive}%
  \BibitemOpen
  \bibfield  {author} {\bibinfo {author} {\bibfnamefont {S.}~\bibnamefont
  {Carrara}}, \bibinfo {author} {\bibfnamefont {D.}~\bibnamefont {Sacchetto}},
  \bibinfo {author} {\bibfnamefont {M.-A.}\ \bibnamefont {Doucey}}, \bibinfo
  {author} {\bibfnamefont {C.}~\bibnamefont {Baj-Rossi}}, \bibinfo {author}
  {\bibfnamefont {G.}~\bibnamefont {De~Micheli}}, \ and\ \bibinfo {author}
  {\bibfnamefont {Y.}~\bibnamefont {Leblebici}},\ }\href@noop {} {\bibfield
  {journal} {\bibinfo  {journal} {Sensors and Actuators B: Chemical}\ }\textbf
  {\bibinfo {volume} {171}},\ \bibinfo {pages} {449} (\bibinfo {year}
  {2012})}\BibitemShut {NoStop}%
\bibitem [{\citenamefont {Sacchetto}\ \emph {et~al.}(2014)\citenamefont
  {Sacchetto}, \citenamefont {Leblebici},\ and\ \citenamefont
  {De~Micheli}}]{Sacchetto14}%
  \BibitemOpen
  \bibfield  {author} {\bibinfo {author} {\bibfnamefont {D.}~\bibnamefont
  {Sacchetto}}, \bibinfo {author} {\bibfnamefont {Y.}~\bibnamefont
  {Leblebici}}, \ and\ \bibinfo {author} {\bibfnamefont {G.}~\bibnamefont
  {De~Micheli}},\ }\enquote {\bibinfo {title} {Memristors and memristive
  systems},}\ \ (\bibinfo  {publisher} {Springer New York},\ \bibinfo {address}
  {New York, NY},\ \bibinfo {year} {2014})\ Chap.\ \bibinfo {chapter} {Silicon
  Nanowire-Based Memristive Devices}, pp.\ \bibinfo {pages}
  {253--280}\BibitemShut {NoStop}%
\bibitem [{\citenamefont {Puppo}\ \emph {et~al.}(2016)\citenamefont {Puppo},
  \citenamefont {Traversa}, \citenamefont {Di~Ventra}, \citenamefont
  {De~Micheli},\ and\ \citenamefont {Carrara}}]{puppo2016surface}%
  \BibitemOpen
  \bibfield  {author} {\bibinfo {author} {\bibfnamefont {F.}~\bibnamefont
  {Puppo}}, \bibinfo {author} {\bibfnamefont {F.~L.}\ \bibnamefont {Traversa}},
  \bibinfo {author} {\bibfnamefont {M.}~\bibnamefont {Di~Ventra}}, \bibinfo
  {author} {\bibfnamefont {G.}~\bibnamefont {De~Micheli}}, \ and\ \bibinfo
  {author} {\bibfnamefont {S.}~\bibnamefont {Carrara}},\ }\href@noop {}
  {\bibfield  {journal} {\bibinfo  {journal} {Nanotechnology}\ }\textbf
  {\bibinfo {volume} {27}},\ \bibinfo {pages} {345503} (\bibinfo {year}
  {2016})}\BibitemShut {NoStop}%
\bibitem [{\citenamefont {Rajeev}\ \emph {et~al.}(2017)\citenamefont {Rajeev},
  \citenamefont {Opoku}, \citenamefont {Stolojan}, \citenamefont
  {Constantinou},\ and\ \citenamefont {Shkunov}}]{rajeev2017effect}%
  \BibitemOpen
  \bibfield  {author} {\bibinfo {author} {\bibfnamefont {K.~P.}\ \bibnamefont
  {Rajeev}}, \bibinfo {author} {\bibfnamefont {C.}~\bibnamefont {Opoku}},
  \bibinfo {author} {\bibfnamefont {V.}~\bibnamefont {Stolojan}}, \bibinfo
  {author} {\bibfnamefont {M.}~\bibnamefont {Constantinou}}, \ and\ \bibinfo
  {author} {\bibfnamefont {M.}~\bibnamefont {Shkunov}},\ }\href@noop {}
  {\bibfield  {journal} {\bibinfo  {journal} {Nanoscience and Nanoengineering}\
  }\textbf {\bibinfo {volume} {5}},\ \bibinfo {pages} {17} (\bibinfo {year}
  {2017})}\BibitemShut {NoStop}%
\bibitem [{\citenamefont {Seki}(2016)}]{seki2016equivalent}%
  \BibitemOpen
  \bibfield  {author} {\bibinfo {author} {\bibfnamefont {K.}~\bibnamefont
  {Seki}},\ }\href@noop {} {\bibfield  {journal} {\bibinfo  {journal} {Applied
  Physics Letters}\ }\textbf {\bibinfo {volume} {109}},\ \bibinfo {pages}
  {033905} (\bibinfo {year} {2016})}\BibitemShut {NoStop}%
\bibitem [{\citenamefont {Chen}\ \emph {et~al.}(2015)\citenamefont {Chen},
  \citenamefont {Yang}, \citenamefont {Zheng}, \citenamefont {Wu},
  \citenamefont {Li}, \citenamefont {Yan}, \citenamefont {Bisquert},
  \citenamefont {Garcia-Belmonte}, \citenamefont {Zhu},\ and\ \citenamefont
  {Priya}}]{chen2015impact}%
  \BibitemOpen
  \bibfield  {author} {\bibinfo {author} {\bibfnamefont {B.}~\bibnamefont
  {Chen}}, \bibinfo {author} {\bibfnamefont {M.}~\bibnamefont {Yang}}, \bibinfo
  {author} {\bibfnamefont {X.}~\bibnamefont {Zheng}}, \bibinfo {author}
  {\bibfnamefont {C.}~\bibnamefont {Wu}}, \bibinfo {author} {\bibfnamefont
  {W.}~\bibnamefont {Li}}, \bibinfo {author} {\bibfnamefont {Y.}~\bibnamefont
  {Yan}}, \bibinfo {author} {\bibfnamefont {J.}~\bibnamefont {Bisquert}},
  \bibinfo {author} {\bibfnamefont {G.}~\bibnamefont {Garcia-Belmonte}},
  \bibinfo {author} {\bibfnamefont {K.}~\bibnamefont {Zhu}}, \ and\ \bibinfo
  {author} {\bibfnamefont {S.}~\bibnamefont {Priya}},\ }\href@noop {}
  {\bibfield  {journal} {\bibinfo  {journal} {The journal of physical chemistry
  letters}\ }\textbf {\bibinfo {volume} {6}},\ \bibinfo {pages} {4693}
  (\bibinfo {year} {2015})}\BibitemShut {NoStop}%
\bibitem [{\citenamefont {Tress}\ \emph {et~al.}(2016)\citenamefont {Tress},
  \citenamefont {Correa~Baena}, \citenamefont {Saliba}, \citenamefont {Abate},\
  and\ \citenamefont {Graetzel}}]{tress2016inverted}%
  \BibitemOpen
  \bibfield  {author} {\bibinfo {author} {\bibfnamefont {W.}~\bibnamefont
  {Tress}}, \bibinfo {author} {\bibfnamefont {J.~P.}\ \bibnamefont
  {Correa~Baena}}, \bibinfo {author} {\bibfnamefont {M.}~\bibnamefont
  {Saliba}}, \bibinfo {author} {\bibfnamefont {A.}~\bibnamefont {Abate}}, \
  and\ \bibinfo {author} {\bibfnamefont {M.}~\bibnamefont {Graetzel}},\
  }\href@noop {} {\bibfield  {journal} {\bibinfo  {journal} {Advanced Energy
  Materials}\ }\textbf {\bibinfo {volume} {6}},\ \bibinfo {pages} {1600396}
  (\bibinfo {year} {2016})}\BibitemShut {NoStop}%
\bibitem [{\citenamefont {Moiz}\ \emph {et~al.}(2020)\citenamefont {Moiz},
  \citenamefont {Alahmadi},\ and\ \citenamefont {Aljohani}}]{moiz2020design}%
  \BibitemOpen
  \bibfield  {author} {\bibinfo {author} {\bibfnamefont {S.~A.}\ \bibnamefont
  {Moiz}}, \bibinfo {author} {\bibfnamefont {A.}~\bibnamefont {Alahmadi}}, \
  and\ \bibinfo {author} {\bibfnamefont {A.~J.}\ \bibnamefont {Aljohani}},\
  }\href@noop {} {\bibfield  {journal} {\bibinfo  {journal} {Energies}\
  }\textbf {\bibinfo {volume} {13}},\ \bibinfo {pages} {3797} (\bibinfo {year}
  {2020})}\BibitemShut {NoStop}%
\bibitem [{\citenamefont {Yu}\ \emph {et~al.}(2016)\citenamefont {Yu},
  \citenamefont {Wu}, \citenamefont {Liu}, \citenamefont {Xiong}, \citenamefont
  {Jagadish},\ and\ \citenamefont {Wang}}]{yu2016design}%
  \BibitemOpen
  \bibfield  {author} {\bibinfo {author} {\bibfnamefont {P.}~\bibnamefont
  {Yu}}, \bibinfo {author} {\bibfnamefont {J.}~\bibnamefont {Wu}}, \bibinfo
  {author} {\bibfnamefont {S.}~\bibnamefont {Liu}}, \bibinfo {author}
  {\bibfnamefont {J.}~\bibnamefont {Xiong}}, \bibinfo {author} {\bibfnamefont
  {C.}~\bibnamefont {Jagadish}}, \ and\ \bibinfo {author} {\bibfnamefont
  {Z.~M.}\ \bibnamefont {Wang}},\ }\href@noop {} {\bibfield  {journal}
  {\bibinfo  {journal} {Nano Today}\ }\textbf {\bibinfo {volume} {11}},\
  \bibinfo {pages} {704} (\bibinfo {year} {2016})}\BibitemShut {NoStop}%
\bibitem [{\citenamefont {Rurali}(2010)}]{rurali2010colloquium}%
  \BibitemOpen
  \bibfield  {author} {\bibinfo {author} {\bibfnamefont {R.}~\bibnamefont
  {Rurali}},\ }\href@noop {} {\bibfield  {journal} {\bibinfo  {journal}
  {Reviews of Modern Physics}\ }\textbf {\bibinfo {volume} {82}},\ \bibinfo
  {pages} {427} (\bibinfo {year} {2010})}\BibitemShut {NoStop}%
\bibitem [{\citenamefont {Fakhri}\ \emph {et~al.}(2021)\citenamefont {Fakhri},
  \citenamefont {Sultan}, \citenamefont {Manolescu}, \citenamefont
  {Ingvarsson}, \citenamefont {Plugaru}, \citenamefont {Plugaru},\ and\
  \citenamefont {Svavarsson}}]{fakhri2021synthesis}%
  \BibitemOpen
  \bibfield  {author} {\bibinfo {author} {\bibfnamefont {E.}~\bibnamefont
  {Fakhri}}, \bibinfo {author} {\bibfnamefont {M.}~\bibnamefont {Sultan}},
  \bibinfo {author} {\bibfnamefont {A.}~\bibnamefont {Manolescu}}, \bibinfo
  {author} {\bibfnamefont {S.}~\bibnamefont {Ingvarsson}}, \bibinfo {author}
  {\bibfnamefont {N.}~\bibnamefont {Plugaru}}, \bibinfo {author} {\bibfnamefont
  {R.}~\bibnamefont {Plugaru}}, \ and\ \bibinfo {author} {\bibfnamefont
  {H.}~\bibnamefont {Svavarsson}},\ }in\ \href@noop {} {\emph {\bibinfo
  {booktitle} {2021 International Semiconductor Conference (CAS)}}}\ (\bibinfo
  {organization} {IEEE},\ \bibinfo {year} {2021})\ pp.\ \bibinfo {pages}
  {147--150}\BibitemShut {NoStop}%
\bibitem [{\citenamefont {Stanchu}\ \emph {et~al.}(2017)\citenamefont
  {Stanchu}, \citenamefont {Kuchuk}, \citenamefont {Barchuk}, \citenamefont
  {Mazur}, \citenamefont {Kladko}, \citenamefont {Wang}, \citenamefont
  {Rafaja},\ and\ \citenamefont {Salamo}}]{stanchu2017asymmetrical}%
  \BibitemOpen
  \bibfield  {author} {\bibinfo {author} {\bibfnamefont {H.}~\bibnamefont
  {Stanchu}}, \bibinfo {author} {\bibfnamefont {A.}~\bibnamefont {Kuchuk}},
  \bibinfo {author} {\bibfnamefont {M.}~\bibnamefont {Barchuk}}, \bibinfo
  {author} {\bibfnamefont {Y.~I.}\ \bibnamefont {Mazur}}, \bibinfo {author}
  {\bibfnamefont {V.}~\bibnamefont {Kladko}}, \bibinfo {author} {\bibfnamefont
  {Z.~M.}\ \bibnamefont {Wang}}, \bibinfo {author} {\bibfnamefont
  {D.}~\bibnamefont {Rafaja}}, \ and\ \bibinfo {author} {\bibfnamefont
  {G.}~\bibnamefont {Salamo}},\ }\href@noop {} {\bibfield  {journal} {\bibinfo
  {journal} {CrystEngComm}\ }\textbf {\bibinfo {volume} {19}},\ \bibinfo
  {pages} {2977} (\bibinfo {year} {2017})}\BibitemShut {NoStop}%
\bibitem [{\citenamefont {Kaganer}\ \emph {et~al.}(2016)\citenamefont
  {Kaganer}, \citenamefont {Jenichen},\ and\ \citenamefont
  {Brandt}}]{kaganer2016elastic}%
  \BibitemOpen
  \bibfield  {author} {\bibinfo {author} {\bibfnamefont {V.~M.}\ \bibnamefont
  {Kaganer}}, \bibinfo {author} {\bibfnamefont {B.}~\bibnamefont {Jenichen}}, \
  and\ \bibinfo {author} {\bibfnamefont {O.}~\bibnamefont {Brandt}},\
  }\href@noop {} {\bibfield  {journal} {\bibinfo  {journal} {Physical Review
  Applied}\ }\textbf {\bibinfo {volume} {6}},\ \bibinfo {pages} {064023}
  (\bibinfo {year} {2016})}\BibitemShut {NoStop}%
\bibitem [{\citenamefont {Romanitan}\ \emph {et~al.}(2019)\citenamefont
  {Romanitan}, \citenamefont {Kusko}, \citenamefont {Popescu}, \citenamefont
  {Varasteanu}, \citenamefont {Radoi},\ and\ \citenamefont
  {Pachiu}}]{romanitan2019unravelling}%
  \BibitemOpen
  \bibfield  {author} {\bibinfo {author} {\bibfnamefont {C.}~\bibnamefont
  {Romanitan}}, \bibinfo {author} {\bibfnamefont {M.}~\bibnamefont {Kusko}},
  \bibinfo {author} {\bibfnamefont {M.}~\bibnamefont {Popescu}}, \bibinfo
  {author} {\bibfnamefont {P.}~\bibnamefont {Varasteanu}}, \bibinfo {author}
  {\bibfnamefont {A.}~\bibnamefont {Radoi}}, \ and\ \bibinfo {author}
  {\bibfnamefont {C.}~\bibnamefont {Pachiu}},\ }\href@noop {} {\bibfield
  {journal} {\bibinfo  {journal} {Journal of Applied Crystallography}\ }\textbf
  {\bibinfo {volume} {52}},\ \bibinfo {pages} {1077} (\bibinfo {year}
  {2019})}\BibitemShut {NoStop}%
\bibitem [{\citenamefont {Congli}\ \emph {et~al.}(2013)\citenamefont {Congli},
  \citenamefont {Hao}, \citenamefont {Huanhuan}, \citenamefont {Jingjing},
  \citenamefont {Yu}, \citenamefont {Yong}, \citenamefont {Zhifeng},\ and\
  \citenamefont {Xiaosong}}]{congli2013synthesis}%
  \BibitemOpen
  \bibfield  {author} {\bibinfo {author} {\bibfnamefont {S.}~\bibnamefont
  {Congli}}, \bibinfo {author} {\bibfnamefont {H.}~\bibnamefont {Hao}},
  \bibinfo {author} {\bibfnamefont {F.}~\bibnamefont {Huanhuan}}, \bibinfo
  {author} {\bibfnamefont {X.}~\bibnamefont {Jingjing}}, \bibinfo {author}
  {\bibfnamefont {C.}~\bibnamefont {Yu}}, \bibinfo {author} {\bibfnamefont
  {J.}~\bibnamefont {Yong}}, \bibinfo {author} {\bibfnamefont {J.}~\bibnamefont
  {Zhifeng}}, \ and\ \bibinfo {author} {\bibfnamefont {S.}~\bibnamefont
  {Xiaosong}},\ }\href@noop {} {\bibfield  {journal} {\bibinfo  {journal}
  {Applied Surface Science}\ }\textbf {\bibinfo {volume} {282}},\ \bibinfo
  {pages} {259} (\bibinfo {year} {2013})}\BibitemShut {NoStop}%
\bibitem [{\citenamefont {Canham}(2020)}]{canham2020introductory}%
  \BibitemOpen
  \bibfield  {author} {\bibinfo {author} {\bibfnamefont {L.}~\bibnamefont
  {Canham}},\ }\href@noop {} {\bibfield  {journal} {\bibinfo  {journal}
  {Faraday Discussions}\ }\textbf {\bibinfo {volume} {222}},\ \bibinfo {pages}
  {10} (\bibinfo {year} {2020})}\BibitemShut {NoStop}%
\bibitem [{\citenamefont {Jung}\ \emph {et~al.}(2019)\citenamefont {Jung},
  \citenamefont {Sohn},\ and\ \citenamefont {Kim}}]{jung2019optical}%
  \BibitemOpen
  \bibfield  {author} {\bibinfo {author} {\bibfnamefont {D.}~\bibnamefont
  {Jung}}, \bibinfo {author} {\bibfnamefont {H.}~\bibnamefont {Sohn}}, \ and\
  \bibinfo {author} {\bibfnamefont {Y.}~\bibnamefont {Kim}},\ }\href@noop {}
  {\bibfield  {journal} {\bibinfo  {journal} {Journal of the Korean Physical
  Society}\ }\textbf {\bibinfo {volume} {74}},\ \bibinfo {pages} {140}
  (\bibinfo {year} {2019})}\BibitemShut {NoStop}%
\bibitem [{\citenamefont {Lin}\ \emph {et~al.}(2010)\citenamefont {Lin},
  \citenamefont {Guo}, \citenamefont {Sun}, \citenamefont {Feng},\ and\
  \citenamefont {Wang}}]{lin2010synthesis}%
  \BibitemOpen
  \bibfield  {author} {\bibinfo {author} {\bibfnamefont {L.}~\bibnamefont
  {Lin}}, \bibinfo {author} {\bibfnamefont {S.}~\bibnamefont {Guo}}, \bibinfo
  {author} {\bibfnamefont {X.}~\bibnamefont {Sun}}, \bibinfo {author}
  {\bibfnamefont {J.}~\bibnamefont {Feng}}, \ and\ \bibinfo {author}
  {\bibfnamefont {Y.}~\bibnamefont {Wang}},\ }\href@noop {} {\bibfield
  {journal} {\bibinfo  {journal} {Nanoscale research letters}\ }\textbf
  {\bibinfo {volume} {5}},\ \bibinfo {pages} {1822} (\bibinfo {year}
  {2010})}\BibitemShut {NoStop}%
\bibitem [{\citenamefont {Razek}\ \emph {et~al.}(2014)\citenamefont {Razek},
  \citenamefont {Swillam},\ and\ \citenamefont {Allam}}]{razek2014vertically}%
  \BibitemOpen
  \bibfield  {author} {\bibinfo {author} {\bibfnamefont {S.~A.}\ \bibnamefont
  {Razek}}, \bibinfo {author} {\bibfnamefont {M.~A.}\ \bibnamefont {Swillam}},
  \ and\ \bibinfo {author} {\bibfnamefont {N.~K.}\ \bibnamefont {Allam}},\
  }\href@noop {} {\bibfield  {journal} {\bibinfo  {journal} {Journal of Applied
  Physics}\ }\textbf {\bibinfo {volume} {115}},\ \bibinfo {pages} {194305}
  (\bibinfo {year} {2014})}\BibitemShut {NoStop}%
\bibitem [{\citenamefont {Naffeti}\ \emph {et~al.}(2020)\citenamefont
  {Naffeti}, \citenamefont {Postigo}, \citenamefont {Chtourou},\ and\
  \citenamefont {Za{\"\i}bi}}]{naffeti2020highly}%
  \BibitemOpen
  \bibfield  {author} {\bibinfo {author} {\bibfnamefont {M.}~\bibnamefont
  {Naffeti}}, \bibinfo {author} {\bibfnamefont {P.~A.}\ \bibnamefont
  {Postigo}}, \bibinfo {author} {\bibfnamefont {R.}~\bibnamefont {Chtourou}}, \
  and\ \bibinfo {author} {\bibfnamefont {M.~A.}\ \bibnamefont {Za{\"\i}bi}},\
  }\href@noop {} {\bibfield  {journal} {\bibinfo  {journal} {Nanomaterials}\
  }\textbf {\bibinfo {volume} {10}},\ \bibinfo {pages} {1434} (\bibinfo {year}
  {2020})}\BibitemShut {NoStop}%
\bibitem [{\citenamefont {Leontis}\ \emph {et~al.}(2013)\citenamefont
  {Leontis}, \citenamefont {Othonos},\ and\ \citenamefont
  {Nassiopoulou}}]{leontis2013structure}%
  \BibitemOpen
  \bibfield  {author} {\bibinfo {author} {\bibfnamefont {I.}~\bibnamefont
  {Leontis}}, \bibinfo {author} {\bibfnamefont {A.}~\bibnamefont {Othonos}}, \
  and\ \bibinfo {author} {\bibfnamefont {A.~G.}\ \bibnamefont {Nassiopoulou}},\
  }\href@noop {} {\bibfield  {journal} {\bibinfo  {journal} {Nanoscale research
  letters}\ }\textbf {\bibinfo {volume} {8}},\ \bibinfo {pages} {1} (\bibinfo
  {year} {2013})}\BibitemShut {NoStop}%
\bibitem [{\citenamefont {Okayama}\ \emph {et~al.}(2009)\citenamefont
  {Okayama}, \citenamefont {Fukami}, \citenamefont {Plugaru}, \citenamefont
  {Sakka},\ and\ \citenamefont {Ogata}}]{okayama2009ordering}%
  \BibitemOpen
  \bibfield  {author} {\bibinfo {author} {\bibfnamefont {H.}~\bibnamefont
  {Okayama}}, \bibinfo {author} {\bibfnamefont {K.}~\bibnamefont {Fukami}},
  \bibinfo {author} {\bibfnamefont {R.}~\bibnamefont {Plugaru}}, \bibinfo
  {author} {\bibfnamefont {T.}~\bibnamefont {Sakka}}, \ and\ \bibinfo {author}
  {\bibfnamefont {Y.~H.}\ \bibnamefont {Ogata}},\ }\href@noop {} {\bibfield
  {journal} {\bibinfo  {journal} {Journal of The Electrochemical Society}\
  }\textbf {\bibinfo {volume} {157}},\ \bibinfo {pages} {D54} (\bibinfo {year}
  {2009})}\BibitemShut {NoStop}%
\bibitem [{\citenamefont {Amri}\ \emph {et~al.}(2020)\citenamefont {Amri},
  \citenamefont {Ezzaouia},\ and\ \citenamefont
  {Ouertani}}]{amri2020photoluminescence}%
  \BibitemOpen
  \bibfield  {author} {\bibinfo {author} {\bibfnamefont {C.}~\bibnamefont
  {Amri}}, \bibinfo {author} {\bibfnamefont {H.}~\bibnamefont {Ezzaouia}}, \
  and\ \bibinfo {author} {\bibfnamefont {R.}~\bibnamefont {Ouertani}},\
  }\href@noop {} {\bibfield  {journal} {\bibinfo  {journal} {Chinese Journal of
  Physics}\ }\textbf {\bibinfo {volume} {63}},\ \bibinfo {pages} {325}
  (\bibinfo {year} {2020})}\BibitemShut {NoStop}%
\bibitem [{\citenamefont {Wang}\ \emph {et~al.}(2012)\citenamefont {Wang},
  \citenamefont {Chao},\ and\ \citenamefont {Su}}]{wang2012electrochemically}%
  \BibitemOpen
  \bibfield  {author} {\bibinfo {author} {\bibfnamefont {R.-C.}\ \bibnamefont
  {Wang}}, \bibinfo {author} {\bibfnamefont {C.-Y.}\ \bibnamefont {Chao}}, \
  and\ \bibinfo {author} {\bibfnamefont {W.-S.}\ \bibnamefont {Su}},\
  }\href@noop {} {\bibfield  {journal} {\bibinfo  {journal} {Acta materialia}\
  }\textbf {\bibinfo {volume} {60}},\ \bibinfo {pages} {2097} (\bibinfo {year}
  {2012})}\BibitemShut {NoStop}%
\bibitem [{\citenamefont {Dawood}\ \emph {et~al.}(2010)\citenamefont {Dawood},
  \citenamefont {Liew}, \citenamefont {Lianto}, \citenamefont {Hong},
  \citenamefont {Tripathy}, \citenamefont {Thong},\ and\ \citenamefont
  {Choi}}]{dawood2010interference}%
  \BibitemOpen
  \bibfield  {author} {\bibinfo {author} {\bibfnamefont {M.}~\bibnamefont
  {Dawood}}, \bibinfo {author} {\bibfnamefont {T.}~\bibnamefont {Liew}},
  \bibinfo {author} {\bibfnamefont {P.}~\bibnamefont {Lianto}}, \bibinfo
  {author} {\bibfnamefont {M.}~\bibnamefont {Hong}}, \bibinfo {author}
  {\bibfnamefont {S.}~\bibnamefont {Tripathy}}, \bibinfo {author}
  {\bibfnamefont {J.}~\bibnamefont {Thong}}, \ and\ \bibinfo {author}
  {\bibfnamefont {W.}~\bibnamefont {Choi}},\ }\href@noop {} {\bibfield
  {journal} {\bibinfo  {journal} {Nanotechnology}\ }\textbf {\bibinfo {volume}
  {21}},\ \bibinfo {pages} {205305} (\bibinfo {year} {2010})}\BibitemShut
  {NoStop}%
\bibitem [{\citenamefont {Sze}\ and\ \citenamefont {Ng}(2006)}]{Sze_Ch3}%
  \BibitemOpen
  \bibfield  {author} {\bibinfo {author} {\bibfnamefont {S.~M.}\ \bibnamefont
  {Sze}}\ and\ \bibinfo {author} {\bibfnamefont {K.~K.}\ \bibnamefont {Ng}},\
  }\enquote {\bibinfo {title} {Physics of semiconductor devices},}\ \ (\bibinfo
   {publisher} {John Wiley \& Sons, Ltd},\ \bibinfo {year} {2006})\
  Chap.~\bibinfo {chapter} {3}, pp.\ \bibinfo {pages} {134--196},\ \bibinfo
  {edition} {3rd}\ ed.\BibitemShut {Stop}%
\bibitem [{\citenamefont {Nemnes}\ \emph {et~al.}(2017)\citenamefont {Nemnes},
  \citenamefont {Besleaga}, \citenamefont {Tomulescu}, \citenamefont
  {Pintilie}, \citenamefont {Pintilie}, \citenamefont {Torfason},\ and\
  \citenamefont {Manolescu}}]{nemnes2017dynamic}%
  \BibitemOpen
  \bibfield  {author} {\bibinfo {author} {\bibfnamefont {G.~A.}\ \bibnamefont
  {Nemnes}}, \bibinfo {author} {\bibfnamefont {C.}~\bibnamefont {Besleaga}},
  \bibinfo {author} {\bibfnamefont {A.~G.}\ \bibnamefont {Tomulescu}}, \bibinfo
  {author} {\bibfnamefont {I.}~\bibnamefont {Pintilie}}, \bibinfo {author}
  {\bibfnamefont {L.}~\bibnamefont {Pintilie}}, \bibinfo {author}
  {\bibfnamefont {K.}~\bibnamefont {Torfason}}, \ and\ \bibinfo {author}
  {\bibfnamefont {A.}~\bibnamefont {Manolescu}},\ }\href@noop {} {\bibfield
  {journal} {\bibinfo  {journal} {Solar Energy Materials and Solar Cells}\
  }\textbf {\bibinfo {volume} {159}},\ \bibinfo {pages} {197} (\bibinfo {year}
  {2017})}\BibitemShut {NoStop}%
\bibitem [{\citenamefont {Thissandier}\ \emph {et~al.}(2012)\citenamefont
  {Thissandier}, \citenamefont {{Le Comte}}, \citenamefont {Crosnier},
  \citenamefont {Gentile}, \citenamefont {Bidan}, \citenamefont {Hadji},
  \citenamefont {Brousse},\ and\ \citenamefont {Sadki}}]{THISSANDIER2012109}%
  \BibitemOpen
  \bibfield  {author} {\bibinfo {author} {\bibfnamefont {F.}~\bibnamefont
  {Thissandier}}, \bibinfo {author} {\bibfnamefont {A.}~\bibnamefont {{Le
  Comte}}}, \bibinfo {author} {\bibfnamefont {O.}~\bibnamefont {Crosnier}},
  \bibinfo {author} {\bibfnamefont {P.}~\bibnamefont {Gentile}}, \bibinfo
  {author} {\bibfnamefont {G.}~\bibnamefont {Bidan}}, \bibinfo {author}
  {\bibfnamefont {E.}~\bibnamefont {Hadji}}, \bibinfo {author} {\bibfnamefont
  {T.}~\bibnamefont {Brousse}}, \ and\ \bibinfo {author} {\bibfnamefont
  {S.}~\bibnamefont {Sadki}},\ }\href@noop {} {\bibfield  {journal} {\bibinfo
  {journal} {Electrochemistry Communications}\ }\textbf {\bibinfo {volume}
  {25}},\ \bibinfo {pages} {109} (\bibinfo {year} {2012})}\BibitemShut
  {NoStop}%
\bibitem [{\citenamefont {Rouis}\ \emph {et~al.}(2021)\citenamefont {Rouis},
  \citenamefont {Hizem}, \citenamefont {Hassen},\ and\ \citenamefont
  {Kalboussi}}]{Rouis2021}%
  \BibitemOpen
  \bibfield  {author} {\bibinfo {author} {\bibfnamefont {A.}~\bibnamefont
  {Rouis}}, \bibinfo {author} {\bibfnamefont {N.}~\bibnamefont {Hizem}},
  \bibinfo {author} {\bibfnamefont {M.}~\bibnamefont {Hassen}}, \ and\ \bibinfo
  {author} {\bibfnamefont {A.}~\bibnamefont {Kalboussi}},\ }\href@noop {}
  {\bibfield  {journal} {\bibinfo  {journal} {Silicon}\ } (\bibinfo {year}
  {2021})}\BibitemShut {NoStop}%
\bibitem [{\citenamefont {Rams}\ \emph {et~al.}(1999)\citenamefont {Rams},
  \citenamefont {Mendez}, \citenamefont {Craciun}, \citenamefont {Plugaru},\
  and\ \citenamefont {Piqueras}}]{rams1999cathodoluminescence}%
  \BibitemOpen
  \bibfield  {author} {\bibinfo {author} {\bibfnamefont {J.}~\bibnamefont
  {Rams}}, \bibinfo {author} {\bibfnamefont {B.}~\bibnamefont {Mendez}},
  \bibinfo {author} {\bibfnamefont {G.}~\bibnamefont {Craciun}}, \bibinfo
  {author} {\bibfnamefont {R.}~\bibnamefont {Plugaru}}, \ and\ \bibinfo
  {author} {\bibfnamefont {J.}~\bibnamefont {Piqueras}},\ }\href@noop {}
  {\bibfield  {journal} {\bibinfo  {journal} {Applied physics letters}\
  }\textbf {\bibinfo {volume} {74}},\ \bibinfo {pages} {1728} (\bibinfo {year}
  {1999})}\BibitemShut {NoStop}%
\bibitem [{\citenamefont {Chen}\ \emph {et~al.}(1994)\citenamefont {Chen},
  \citenamefont {Lee},\ and\ \citenamefont {Bosman}}]{chen1994electrical}%
  \BibitemOpen
  \bibfield  {author} {\bibinfo {author} {\bibfnamefont {Z.}~\bibnamefont
  {Chen}}, \bibinfo {author} {\bibfnamefont {T.-Y.}\ \bibnamefont {Lee}}, \
  and\ \bibinfo {author} {\bibfnamefont {G.}~\bibnamefont {Bosman}},\
  }\href@noop {} {\bibfield  {journal} {\bibinfo  {journal} {Applied physics
  letters}\ }\textbf {\bibinfo {volume} {64}},\ \bibinfo {pages} {3446}
  (\bibinfo {year} {1994})}\BibitemShut {NoStop}%
\bibitem [{\citenamefont {Van~Buuren}\ \emph {et~al.}(1993)\citenamefont
  {Van~Buuren}, \citenamefont {Tiedje}, \citenamefont {Dahn},\ and\
  \citenamefont {Way}}]{van1993photoelectron}%
  \BibitemOpen
  \bibfield  {author} {\bibinfo {author} {\bibfnamefont {T.}~\bibnamefont
  {Van~Buuren}}, \bibinfo {author} {\bibfnamefont {T.}~\bibnamefont {Tiedje}},
  \bibinfo {author} {\bibfnamefont {J.}~\bibnamefont {Dahn}}, \ and\ \bibinfo
  {author} {\bibfnamefont {B.}~\bibnamefont {Way}},\ }\href@noop {} {\bibfield
  {journal} {\bibinfo  {journal} {Applied physics letters}\ }\textbf {\bibinfo
  {volume} {63}},\ \bibinfo {pages} {2911} (\bibinfo {year}
  {1993})}\BibitemShut {NoStop}%
\end{thebibliography}%

\end{document}